\documentclass[twocolumn]{aastex631}

\graphicspath{{./}{figures/}}
\usepackage{amsmath}

\usepackage{tensor}

\usepackage{CJK}

\newcommand{\bflux}{\texttt{bflux0}}
\newcommand{\bfluxc}{\texttt{bflux-const}}

\shorttitle{Relativistic Jet from Horizon to Galactic Scales}
\shortauthors{Cho et al.}

\begin{document}
\begin{CJK}{UTF8}{mj}
\title{Bridging Scales in Black Hole Accretion and Feedback: Relativistic Jet linking the Horizon to the Host Galaxy}

\author[0000-0002-2858-9481]{Hyerin Cho (조혜린)}
\affiliation{Center for Astrophysics $\vert$ Harvard \& Smithsonian, 60 Garden Street, Cambridge, MA 02138, USA}
\affiliation{Black Hole Initiative at Harvard University, 20 Garden Street, Cambridge, MA 02138, USA}

\author[0000-0002-0393-7734]{Ben S. Prather} 
\affiliation{CCS-2, Los Alamos National Laboratory, PO Box 1663, Los Alamos, NM 87545, USA}

\author[0000-0002-1919-2730]{Ramesh Narayan}
\affiliation{Center for Astrophysics $\vert$ Harvard \& Smithsonian, 60 Garden Street, Cambridge, MA 02138, USA}
\affiliation{Black Hole Initiative at Harvard University, 20 Garden Street, Cambridge, MA 02138, USA}

\author[0000-0003-1598-0083]{Kung-Yi Su}
\affiliation{Center for Astrophysics $\vert$ Harvard \& Smithsonian, 60 Garden Street, Cambridge, MA 02138, USA}
\affiliation{Black Hole Initiative at Harvard University, 20 Garden Street, Cambridge, MA 02138, USA}

\author[0000-0002-5554-8896]{Priyamvada Natarajan}
\affiliation{Black Hole Initiative at Harvard University, 20 Garden Street, Cambridge, MA 02138, USA}
\affiliation{Department of Astronomy, Yale University, Kline Tower, 266 Whitney Avenue, New Haven, CT 06511, USA}
\affiliation{Department of Physics, Yale University, P.O. Box 208121, New Haven, CT 06520, USA}

\begin{abstract} \label{abstract}
Simulating black hole (BH) accretion and feedback from the horizon to galactic scales is extremely challenging, as it involves a vast range of scales. Recently, our multizone method has successfully achieved global dynamical steady-states of hot accretion flows in three-dimensional general relativistic magnetohydrodynamic (GRMHD) simulations by tracking the bidirectional interaction between a non-spinning BH and its host galaxy. In this paper, we present technical improvements to the method and apply it to spin $a_*=0.9$ BHs, which power relativistic jets. We first test the new multizone set-up with a smaller Bondi radius, $R_B\approx400\,r_g$, where $r_g$ is the gravitational radius. The strongly magnetized accretion launches a relativistic jet with an intermediate feedback efficiency $\eta\sim30\,\%$, in between that of a prograde ($\eta\sim100\,\%$) and retrograde ($\eta\sim 10\,\%$) torus. Interestingly, both prograde and retrograde simulations also eventually converge to the same intermediate efficiency when evolved long enough, as accumulated magnetic fields remove gas rotation. We then extend strongly magnetized simulations to larger Bondi radii, $R_B\approx 2\times10^3,~2\times 10^4,~2\times 10^5\,r_g$. We find that the BH accretion rate $\dot{M}$ is suppressed with respect to the Bondi rate $\dot{M}_B$  as $\dot{M}/\dot{M}_B\propto R_B^{-1/2}$. However, despite some variability, the time-averaged feedback efficiency is $\eta\sim30\,\%$, independent of $R_B$. This suggests that BH feedback efficiency in hot accretion flows is mainly governed by the BH spin ($a_*$) rather than by the galactic properties ($R_B$). From these first-principles simulations, we provide a feedback subgrid prescription for cosmological simulations: $\dot{E}_{\rm fb}=2\times10^{-3}[R_B/(2\times10^5\,r_g)]^{-1/2}\dot{M}_Bc^2$ for BH spin $a_*=0.9$.
\end{abstract}

\section{Introduction} \label{sec:intro}

The masses of central supermassive black holes (SMBHs) and properties of their host galaxies, such as their stellar bulge masses, are tightly correlated despite their disparate scales \citep[e.g., ][]{Magorrian1998, Ferrarese2000, Gebhardt2000, Kormendy2013}. These correlations suggest that the black hole (BH) and the galaxy co-evolve by regulating each other's growth over cosmic time, likely via a continual self-regulation of gas supply to the horizon for BH accretion and star formation in the host galaxy \citep{Alexander+Hickox2012}. 

This co-evolution can be explained by some two-way exchange of information between the small scales relevant to the BH and larger galactic scales, where the relevant lengthscales are, e.g., for M87, $\approx 0.3 \,{\rm mpc}$ and $\gtrsim 100\,{\rm kpc}$, respectively. The coupling requires bi-directional communication that is believed to be mediated by the inflow of gas and energetic outflows powered by the BH. Understanding how exactly the galaxy and its central BH are coupled across up to $9$ orders of magnitudes in spatial scales is an important open question at present.

It has become increasingly clear that relativistic jets are the main feedback mechanism for low-Eddington accretors \citep[e.g., ][]{Fabian2012, Heckman2014} that comprise the majority of nearby active galactic nuclei (AGNs) \citep{Ho2009}. Powerful jets are observed on AU to Mpc scales \citep[see][for a review]{Blandford2019}, which can reach far beyond the visible extent of the host galaxy. With their extended reach, it is believed that jets deposit energy and momentum into the galactic medium and hence maintain the galaxy's global heating/cooling equilibrium \citep{McNamara2007,McNamara2012}.

The state-of-the-art theoretical understanding of low-Eddington accretion, also known as hot accretion, requires a comprehensive description of various factors, including, but not limited to, magnetic fields, turbulence, plasma physics, 3-dimensionality, gravity, and their interplay \citep[see e.g.,][for reviews]{Yuan2014,Davis2020,Komissarov2021}. In the case of magnetic fields, it was recognized long ago that any frozen-in field in the external medium that is advected towards the BH will quickly dominate (energy density $\propto r^{-4}$) as it approaches the BH, strongly impacting the fluid dynamics \citep{Shvartsman1971}. This effect was later confirmed in 3D simulations by \citet{Igumenshchev2003}.
Moreover, they showed that the axisymmetry is strongly broken by the Rayleigh-Taylor (or ``interchange'') instability. The accretion then proceeds in non-axisymmetric narrow streams, which calls for the full consideration of all 3 spatial dimensions. This mode of accretion is referred to as a magnetically arrested disk (MAD) (\citealt{Narayan2003}, see also \citealt{Bisnovatyi-Kogan1974,Bisnovatyi1976}) and has recently been shown to be consistent with the observations of the Event Horizon Telescope (EHT) of M87 * \citep{EHTC2021} and Sgr A*\citep{EHTC2022}. 

A full general relativistic treatment is required to self-consistently simulate the launching of \citet[][BZ hereafter]{Blandford1977}-type jets, caused by the frame-dragging of spacetime \citep{Tchekhovskoy2010,Tchekhovskoy2011} around spinning BHs. Although analytical approaches have provided valuable insights on hot accretion flows, they require many simplifying assumptions (e.g., \citealt{Narayan1994,Narayan2000,Quataert2000}, see \citealt{Yuan2014} for a review). Numerical simulations offer an excellent opportunity to take all the complicated physical processes into account when tackling this complex non-linear problem.

However, simulations face a serious issue---the ``timestep problem''---when dealing with the vastly different length- and timescales that need to be simultaneously probed. The Courant-Friedrichs-Lewy (CFL) condition for numerical stability requires the global timestep of the simulation to be the smallest timescale, typically limited by cells near the event horizon. Since the free-fall timescale scales as $\propto r^{3/2}$, if the outer radii are 6 orders of magnitude larger than the radius $r_H$ of the event horizon, the corresponding characteristic timescale is 9 orders of magnitude longer. Therefore, the global timestep set near the horizon is infinitesimally small compared to any meaningful galactic evolution timescale one wishes to track. 
Even an optimistic back-of-the-envelope calculation indicates an uncomfortably long computational wall time of $\gtrsim 50$ years for a simulation that fully resolves the horizon to converge near a realistic galaxy's Bondi radius, $R_B\sim 10^5\, r_g$, where $r_g$ is the gravitational radius (Equation~\ref{eq:rgtg}).\footnote{This estimate is arrived at for a moderate resolution ($64^2$ in $\theta,\varphi$) and assuming uninterrupted access to 4 GPUs.} The timestep problem is the major obstacle to numerically modeling the interplay between BH and galactic scales.

The simulation community has tackled this obstacle by partitioning the problem into two ranges of scales and studying each independently. Galaxy simulators cover the large galactic scales, and general relativistic magnetohydrodynamic (GRMHD) simulators cover the small scale BH horizon scales. Even with such a partitioning of scales, the two simulations have little to no overlap in spatial scales, as explained below.

Cosmological simulations, including zoom-in galaxy simulations, have successfully reproduced realistic galactic environments \cite[e.g.,][]{Sijacki2015,Rosas-Guevara2016,Weinberger2018,Ricarte2019,Ni2022,Wellons2023} but lack the resolution in the vicinity of the BH (at best $\sim 10\,{\rm pc}$ even for isolated zoom-in galaxy simulations whereas the BH horizon lies at $\lesssim \,\rm{mpc}$ for M87*). Thus, general-relativistic effects are neglected, and the accretion onto and feedback from the BH cannot be tracked at relevant scales. These larger scale simulations address this problem by adopting different subgrid prescriptions to model the unresolved BH accretion and feedback \cite[][etc.]{Li2014,Weinberger2017,Tremmel2017,Fiacconi2018,Angles-Alcazar2021,Talbot2021,Su2021,Koudmani2024,Rennehan2024,Grete2025}. However, the results of these larger scale simulations are highly sensitive to the specific choices made for the BH subgrid prescriptions \citep[e.g.,][]{Wellons2023,Weinberger2023}. This great uncertainty in the modeling of BH physics limits our understanding of BH feedback in the galactic context \citep[see][for reviews]{Somerville2015,Naab2017,Vogelsberger2020,Crain2023}.

GRMHD simulations, on the other hand, self-consistently model BH accretion and feedback from first principles by solving general relativistic equations at high resolutions on the BH horizon scales \citep[e.g.,][]{Komissarov1999,Gammie2003, Tchekhovskoy2011, Porth:2019,Narayan2022,Chatterjee:2023}.
However, the computations are highly expensive and thus convergence is typically reached only on scales up to $\sim 100\,r_g$ (or $r\lesssim 30\,{\rm mpc}$ in physical units for M87*). Since these scales are well separated from galactic scales and the dynamics in this regime is governed entirely by the BH's gravitational field, GRMHD simulations typically resort to an idealized set-up that is far from realistic in terms of the environment in the galactic nucleus. Specifically, the \citet{Fishbone1976} torus model is conventionally adopted as idealized initial conditions.

Recent years have seen significant progress in connecting the BH and galactic scales. Four approaches have been tried. The first is the zoom-in technique, which restarts simulations with higher resolution from converged lower resolution large-scale simulations \citep{Hopkins2010,Ressler2020b,Guo2023,Guo2024,Kaaz2025}. Second, the Lagrangian hyperrefinement for Lagrangian codes adaptively refines the gas particles as they accrete towards the BH \citep{Angles-Alcazar2021, Hopkins2024,Hopkins2024b,Hopkins+2025ISCO}. Even though the above two approaches have successfully captured the accretion from galactic to event horizon scales, their simulations are still subject to the critical timestep issue when high resolution is achieved. Therefore, those two approaches cannot track the back-reaction on large scales due to the feedback from small scales, which limits their capability to study the ongoing bidirectional interaction between the BH and its host galaxy. The third approach is to run GRMHD simulations for an extended duration to gradually reach convergence over a wider dynamic range (\citealt{Lalakos2022,Kaaz2023,Lalakos2024,Galishnikova2025,Lalakos2025}). This self-consistently tracks both accretion and feedback, but because of the timestep issue, it can only be applied over a limited range of scales. These studies artificially reduce the Bondi radius from its true value of $10^5-10^6r_g$ (e.g., Sgr A \citealt{Baganoff2003}, NGC 3115 \citealt{Wong2014}, NGC 1600 \citealt{Runge2021}, M87 \citealt{Dimatteo2003,Russell2018}, M84 \citealt{Bambic2023}, PKS0745-191 \citealt{Hlavacek-Larrondo2025}) to $R_B\approx 100-3000 \,r_g$.\footnote{\citet{Lalakos2025} have recently extended the Bondi radius to $10^4\,r_g$ but the run hasn't yet reached a stable MAD state due to the relatively short evolution of $\lesssim 2\,t_B$, where $t_B$ is the Bondi timescale (Equation~\ref{eq:tB}).}. The latest technical advances, such as external and internal static mesh refinements and local adaptive time stepping \citep{Liska2022}, help speed up simulations to reach convergence up to larger distances \citep{Chatterjee2019}. However, the speed-up is typically only by a factor of a few, which is far short of what is needed to connect SMBH horizon scales to galactic $\sim$kpc scales.

Finally, the fourth methodology, the multizone method \citep{Cho2023,Cho2024}, has opened up a new approach to ``bridge scales'' by tracking both accretion and feedback across the vast dynamic range between the BH and the galaxy. \citet{Cho2024}, in particular, have achieved steady states over an unprecedented range of scales, with Bondi radius as large as $R_B\approx 10^7\,r_g$. The key advantage of the multizone method is that it avoids the timestep issue by partitioning the vast dynamic range into annuli, each of which covers a limited range in radius. This implementation is summarized in Section~\ref{sec:numerical_method} of this paper and is described in full detail in \citet{Cho2024}. As with any method, it is subject to some caveats, as described in Section \ref{sec:caveat}.

\citet{Cho2023,Cho2024} demonstrate that, in the case of purely hydrodynamic spherical accretion on a non-spinning BH, analytic general relativistic Bondi solutions \citep{Michel1972,Shapiro1983} are perfectly reproduced by the multizone method. For strongly magnetized Bondi accretion, the accretion rate $\dot{M}$ is suppressed depending on the Bondi radius $R_B$, scaling as $\dot{M}\propto R_B^{-1/2}$. Also, there is mild energy feedback at the level of $\approx 0.02\dot{M}c^2$, even for non-spinning BHs, powered by reconnection-driven convection. Other findings include a density profile varying as $\rho\propto r^{-1}$ \citep[consistent with other works][]{Ressler2020b,Ressler2021, Chatterjee2022, Guo2023, Guo2024}, and plasma-$\beta$ parameter (the gas to magnetic pressure ratio) saturating at order unity indicative of an extended MAD state \citep[see also][]{Ressler2020a}. Recently, \citet{Guo2025} adopted a similar approach to the multizone method but using traditional static mesh refinement, which solves the timestep problem in a different way. We discuss the methodological similarities and differences with our approach in Section \ref{sec:comparison_other_works}.

The fourth method described above paves the way for developing a physically informed AGN feedback model applicable on galaxy scales. Building on the method's results, \cite{Su2025} implemented the suppressed accretion rate and mild feedback efficiency from \citet{Cho2023,Cho2024} into an isolated galaxy simulation with the same M87* initial conditions as in \cite{Cho2024}. This work represents the first galaxy-scale simulation in which AGN feedback is both physically and self-consistently informed from a GRMHD simulation. While \citet{Su2025} found that the mild feedback power of $0.02\dot{M}c^2$ estimated by \citet{Cho2023,Cho2024} for a non-spinning BH can largely prevent excessive black hole growth, significantly stronger feedback $\gtrsim 0.15 \dot{M} c^2$ is likely required to regulate the black hole accretion rate to the level observed by the EHT. Achieving this larger level of feedback will require a non-zero BH spin, and a GRMHD-based prediction for such a model is therefore critically needed. Improving the multizone methodology to simulate BH feedback for spinning BHs over a wide range of scales is the main focus of this work. Additional ongoing efforts include applying the self-consistent AGN feedback model - developed in \citet{Cho2023,Cho2024} and this work - in cosmological simulations (Su et al. 2025, in prep.) and semi-analytic models (SAMs) (Porras et al. 2025, in prep.) to explore its implications for galaxy-BH coevolution in a cosmological context.

In the present work, we extend and apply the multizone method of \citet{Cho2023,Cho2024} to \citet{Kerr1963} BHs, focusing on a spin value of $a_*=0.9$. A significant difference from our previous work \citep{Cho2023,Cho2024} is that, with a spinning BH, we expect the launching of powerful relativistic jets, which will challenge the numerical scheme. 

Here we adopt units of length and time in terms of the gravitational radius $r_g$ and gravitational time $t_g$, defined as 
\begin{equation}\label{eq:rgtg}
    r_g \equiv \frac{GM_\bullet}{c ^2}, \qquad  t_g \equiv \frac{r_g}{c},
\end{equation}
where $M_\bullet$ is the BH mass.
The Bondi radius is given by,
\begin{equation}\label{eq:rB}
    R_B \equiv \frac{GM_\bullet}{c_{s,\infty} ^2},
\end{equation}
where $c_{s,\infty}=\sqrt{\gamma_{\rm ad}T_{\infty}}$ is the sound speed at infinity, $\gamma_{\rm ad}=5/3$ is the adiabatic index, and $T_\infty$ is the temperature of the external medium outside the Bondi radius in relativistic units. The Bondi timescale $t_B$ is then defined as the free-fall timescale at the Bondi radius,
\begin{equation}\label{eq:tB}
    t_B\equiv \left( \frac{R_B}{r_g} \right)^{3/2}\,t_g.
\end{equation}

The outline of the paper is as follows. In Section~\ref{sec:numerical_method}, we describe the multizone method along with the modifications we have made from our previous set-up in order to handle spinning BHs. In Section~\ref{sec:small_scale}, we apply the new multizone set-up to systems with an artificially small Bondi radius, $R_B\approx 400\,r_g$, where the method can be tested against traditional GRMHD simulations. Then in Section~\ref{sec:large_scale} we vary the Bondi radius in the range $R_B\approx 400-2\times 10^5\,r_g$, the largest size matching what we expect for a realistic system. After presenting the multizone results for spinning BHs, in Section~\ref{sec:discussion}, we discuss the limitations of our multizone method, compare it with the other bridging scales approaches noted above, and provide feedback subgrid prescriptions that can be directly implemented in cosmological simulations. In Section~\ref{sec:conclude}, we present our conclusions and outline future prospects.

\section{Numerical Methods}\label{sec:numerical_method}

We employ the open-source GRMHD code KHARMA\footnote{https://github.com/AFD-Illinois/kharma} (``Kokkos-based High-Accuracy Relativistic Magnetohydrodynamics with Adaptive mesh refinement,'' \citealt{Prather2024}).  KHARMA leverages the Kokkos programming model \citep{trott2021} and Parthenon AMR framework \citep{grete2022} to perform well on diverse CPU and GPU architectures. KHARMA's high performance, especially on GPUs, is essential to running our moderate resolution but very long simulations on reasonable timeframes.

KHARMA solves the equations of ideal GRMHD using a scheme similar to HARM \citep{Gammie2003}, with a second-order finite-volume shock-capturing scheme which conserves rest mass and the MHD stress-energy tensor, $T^{\mu\nu}=\left( \rho + u + p_g + b^2 \right) u^{\mu} u^{\nu} + \left( p_g + b^2/2 \right) g^{\mu \nu} - b^{\mu}b^{\nu}$, where $\rho$ is the rest mass density, $u$ is the internal energy density, $u^\mu$ is the fluid four-velocity, $b^\mu$ is the magnetic four-vector, $p_g=(\gamma_{\rm ad}-1)u$ is the gas pressure, and $g^{\mu\nu}$ is the inverse metric.  The temperature $T$ is in relativistic units such that $p_g=\rho T$.  We refer the reader to \citet{Anile1989,Komissarov1999,Gammie2003} for a primer on GRMHD.  In contrast to most schemes based on HARM, KHARMA employs face-resident magnetic field representation with constrained transport (using the scheme from \citealt{gardiner2005}), allowing for static and adaptive mesh refinement.  In addition, KHARMA employs some stability features such as the first-order flux corrections described in \citet{Beckwith2011}. Use of KHARMA features for this work is described in detail later in this section.

\subsection{Overview of the Multizone Method}

The multizone method \citep{Cho2023,Cho2024} is an efficient tool for finding dynamical steady states on a wide range of spatial and temporal scales. This method overcomes the timestep issue discussed in Section~\ref{sec:intro} by allowing each scale to evolve according to its own characteristic timescale, rather than being tied just to the horizon timescale. By doing so, each radius is given plenty of time to respond to any new information transferred from either smaller or larger scales, thus efficiently reaching a global dynamical equilibrium covering many decades in radius. 

\begin{figure}[]
\gridline{
\fig{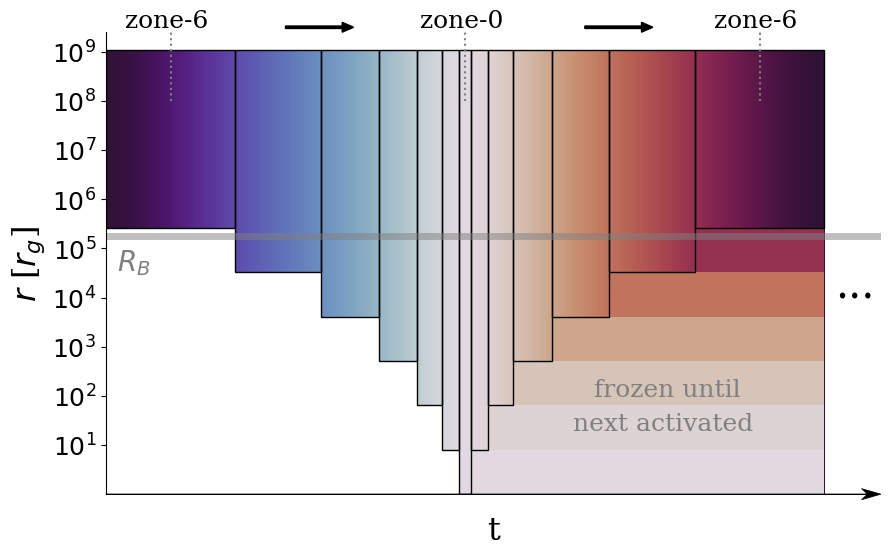}{0.48\textwidth}{}
            }   
\caption{Schematics of the multizone method for spinning BH runs with a Bondi radius $R_B\approx 2\times10^5\,r_g$. The x-axis shows runtime (not to scale) and the y-axis shows radius. Each black box's vertical extent marks the range of active radii for that zone. Changing colors in the horizontal direction within boxes represent time evolution, while fixed colors outside boxes represent the frozen state of the zone while it is inactive. The figure corresponds to one ``V-cycle'' and a complete simulation consists of a large number of these cycles run one after another.
}\label{fig:schematic}
\end{figure}

\subsubsection{Zone Set-Up}
The simulation progresses in a manner analogous to a ``V-cycle'' iterating over zones, $i=0,...,n-1$, where zone-$i$ is an annulus extending from an inner radius $r_{i, \rm in}$ to a common outer radius $r_{\rm out}$. First, the outermost zone $i=n-1$ is simulated while the remaining domains ($r<r_{(n-1), \rm in}$) are held fixed. Then, the zone with the next largest zone becomes active and all regions $r>r_{i, \rm in}$ are simulated. Shifting the active zone continues to zones with smaller $r_{i, \rm in}$ until zone-$0$ with the smallest $r_{0, \rm in}=1\,r_g$ is active. At this stage, the whole simulation domain is evolving and the gas finally reaches the BH horizon. Then, the active zone is switched back out to zones of larger $r_{i, \rm in}$, reversing the sequence. One V-cycle is completed when the zone with the largest $r_{i, \rm in}$ is active again. Within this V-cycle, the inward moving branch of the V-cycle is able to capture the accretion from large to small scales and the outward moving branch captures the feedback from the BH horizon to the galaxy. These V-cycles are repeated hundreds of times until there is sufficient communication between small- and large-scales to achieve a dynamical equilibrium over the entire range of the simulation. Figure~\ref{fig:schematic} illustrates the schematics of the multizone method for a single V-cycle. Essentially, in the multizone method, timesteps vary at every instance when switching between active zones with different $r_{i, \rm in}$. By doing so, even the largest scales evolve over a duration comparable to their long characteristic timescale.

Previously, for non-spinning BH simulations \citep{Cho2023,Cho2024}, we set annuli to have a fixed logarithmic radial width such that both inner $r_{i, \rm in}$ and outer $r_{i, \rm out}$ radii were varied in tandem. In the case of spinning BHs, rapidly propagating relativistic jets can quickly reach a zone's outer boundary $r_{i, \rm out}$ in such a set-up and continuously be subject to the outer boundary condition of the annulus, which can cause problems. Therefore in our current work on spinning BHs, we fix all outer radii $r_{i, \rm out}$ to be equal to $r_{\rm out}$, the largest radius of the simulation. Only the inner radii of zones are varied as $r_{i, \rm in} = 8^i\,r_g$. See zone details in \autoref{tab:multizone_setup} for the case of a run with Bondi radius $R_B\approx 2\times 10^5\,r_g$ and $r_{\rm out}\approx 10^9\,r_g$ (model \texttt{2e5} in \autoref{tab:run_summary}). Here we have chosen a base of $b=8$ (see \citealt{Cho2024} for tests of the effect of using different base values $b$).

\begin{table}[h!]
\centering
 \begin{tabular}{c c c} 
 \hline
Zone \# $i$ & $r_{i,\rm in}$ $[r_g]$ & $r_{\rm out}$ $[r_g]$ \\ [0.5ex] 
 \hline\hline
  0 & $8^{0}=1$ &  \\
  1 & $8^{1}$ &  \\
  \vdots & \vdots & \vdots \\
  i & $8^i$ & $\approx 10^9$ \\
  \vdots & \vdots & \vdots \\
  6 & $8^6$ &  \\ [1ex]
 \hline
 \end{tabular}
 \caption{The multizone set-up using $n=7$ zones for a spinning BH simulation with $R_B\approx 2\times 10^5\,r_g$ (model \texttt{2e5} in \autoref{tab:run_summary}). In contrast to the set-up for non-spinning BHs \citep{Cho2023,Cho2024}, here the outer radii of all the zones are kept fixed at the largest radius of the simulation. This avoids problematic boundary effects as the relativistic jet propagates outward.
 \label{tab:multizone_setup}}
\end{table}

\subsubsection{Boundary Conditions}\label{sec:boundary_conditions}

At the innermost radius $r_{0,\rm in}$, the gas flows inwards into the BH and at the outermost radius $r_{\rm out}$, gas can freely flow out of the simulation domain. Both correspond to ``outflow'' radial boundary conditions, a conventional choice for GRMHD simulations. Reflecting and periodic boundary conditions are used for $\theta$ and $\varphi$, respectively.

The multizone scheme introduces extra internal radial boundaries at $r_{i>0, \rm in}$ as shown in Table~\ref{tab:multizone_setup}. At all these internal boundaries we apply Dirichlet boundary conditions, where $\rho, u, u^\mu, b^\mu$ are fixed at frozen values for the duration of the active zone's runtime.

The magnetic fields require an additional treatment at the radial Dirichlet boundaries in order to preserve the magnetic divergence $\nabla\cdot B=0$. This is because the magnetic fields are frozen at the internal boundaries while the adjacent active regions are evolving the magnetic field. As a result, a non-zero divergence will quickly develop if no prescriptions are adopted. In \citet{Cho2024}, two prescriptions, \bflux{} and \bfluxc{}, were introduced. Simply put, \bflux{} anchors the magnetic fields at the $r_{i, \rm in}$ boundary, which can significantly stretch out the field lines and create a strong stress. On the other hand, \bfluxc{} allows coherent sliding of field lines with respect to the internal boundary in an axi-symmetric manner. See Appendix E of \citet{Cho2024} for tests and comparisons. \citet{Cho2024} showed that non-spinning BH simulations are insensitive to the choice of magnetic field prescription at the radial Dirichlet boundaries due to a lack of coherent rotation of gas and field lines. However, as will be demonstrated in Section~\ref{sec:bc_comparison}, the \bfluxc{} prescription better captures the physics of the relativistic jets in the case of the spinning BHs. This is not surprising because the BZ mechanism operates by coherently rotating magnetic field lines via frame-dragging. Therefore in this work, we apply the \bfluxc{} prescription at all internal boundaries. The implementation used here is described in \autoref{sec:magnetic_boundary_conditions}.

\subsubsection{Run Duration for Each Zone}

The run duration for each zone for these spinning BH runs is modified compared to our previous non-spinning BH simulations. In \citet{Cho2024} the runtime for zone-$i$ was determined by an informed guess of the characteristic timescale, $t_{\rm char}\equiv r/ \sqrt{v_{\rm ff}^2 + c_{s,\infty}^2 }$, where $v_{\rm ff} = \sqrt{GM_\bullet/r}$ is the free-fall velocity. This is a reasonable choice in the absence of jets because the relevant timescales for the dynamics are comparable to the free-fall timescale at $r<R_B$ and the sound-crossing timescale at $r>R_B$. However, the presence of relativistic jets can significantly shorten the characteristic timescale at each radius.

Therefore, instead of using a pre-determined runtime per zone, we evolve for a time $8000\,\Delta t$ for each active zone, where the timestep $\Delta t$ is determined on-the-fly following the CFL condition for each zone. By doing so, we guarantee that each zone is evolved according to its instantaneous characteristic timescale, which is usually set by the jet. This can also prevent a severe accumulation of magnetic tension in the case where the timescales are tremendously shortened and hence numerous steps should be taken until the pre-determined runtime is completed. As with the non-spinning BH set-up, we run zone-$0$ for 10 times longer ($80000\,\Delta t$) to allow extra relaxation of the whole simulation domain. Since there are no internal boundaries for zone-$0$, we can run this zone as long as we wish without any fear of artificial stresses developing.

\subsection{Coordinates and Polar Static Mesh Refinement}\label{sec:smr}
In spherical coordinates, the timesteps are commonly limited by the tiny azimuthal cell length at the poles ($\theta = 0,\pi$) due to the coordinate singularity. Since the azimuthal length $l_\varphi$ of the polar cells is proportional to $l_\varphi\propto \sin(\Delta \theta)\Delta \varphi$, the non-spinning BH simulations \citep{Cho2023, Cho2024} addressed this problem by lowering $\theta$ resolution near the poles using a version of the funky modified Kerr-Schild \citep[FMKS, e.g., ][]{Wong2021} or the wide-pole Kerr-Schild \citep[WKS, ][]{Cho2024} coordinate systems. Lowering the polar resolution could still decently capture quasi-spherical accretion and feedback for non-spinning BH problems.

In the current work, where we  focus on relativistic jets, keeping a high $\theta$ resolution near the poles is critical to capture jet physics. However, a high polar resolution can also significantly slow down the simulation due to constrained timesteps. We handle this problem with an internal static mesh refinement (ISMR) technique that coarsens the azimuthal $\varphi$ resolution near the poles while maintaining the original radial $r$ and polar $\theta$ resolutions, similar to the method of \citet{Liska2022}. Detailed implementation of the ISMR is explained in Appendix~\ref{sec:smr_appendix}. We apply 4 levels of ISMR for the 4 layers of $\theta$ adjacent to the poles for all runs.

With the inclusion of the ISMR technique, keeping a high $\theta$ resolution near the poles is now possible without paying the price of extremely small timesteps. In our work, we employ an exponential Kerr-Schild (EKS) coordinate system where the grid spacing is uniform in $\log(r)$, $\theta$, $\varphi$. In some cases, we use a jet Kerr-Schild (JKS) coordinate which even further enhances resolution near the poles, following the collimating jet morphology. The JKS coordinate system is described in Appendix~\ref{sec:JKS}.

\subsection{Additional Numerical Set-up Details}

Our fiducial initial condition (IC) is the strongly magnetized Bondi solution, identical to the runs in \citet{Cho2023,Cho2024}. However, as we also explore two additional ICs for the reduced scale separation problem, we describe all three types of ICs in Section~\ref{sec:initial_condition}.

The simulations use face-centered magnetic fields which are evolved using the constrained transport algorithm \citep{Balsara1999,gardiner2005}.
At the polar reflecting boundaries in spherical coordinates, small-scale toroidal field loops can collect and grow, leading to unrealistic or disruptive behavior, in contrast to a realistic flow where these field loops will eventually shrink and reconnect. We mimic the expected reconnection by axisymmetrically subtracting the azimuthally-averaged toroidal field $\left<B^\varphi_{\rm pole}\right>_\varphi$ at each timestep only at the first layer of polar cells: $B_{\rm pole}^\varphi(r, \varphi) \to B_{\rm pole}^\varphi(r, \varphi) - \left<B_{\rm pole}^\varphi\right>_\varphi(r)$. Here, $B^i$ in capital letters are magnetic field 3-vectors. In this ``reconnection'' process, the conserved variables remain unchanged, so the total energy is conserved.

For numerical stability, simulation floors and ceilings are applied to ensure the following conditions: density $\rho > 10^{-6} \,r^{-3/2}$, internal energy $u > 10^{-8}\,r^{-5/2}$, magnetic-to-internal energy ratio $b^2/u < 10^4$, and magnetization $b^2/\rho < 100$. The Lorentz factor measured in the Eulerian frame is kept below $\Gamma<\Gamma_{\rm max} = 10$.
First-order flux corrections \citep{Beckwith2011} are applied to any cell which would fall outside these limits, and if the first-order value still needs correction, floor material and energy are added in the Eulerian (normal) observer frame \citep{McKinney2012}. In the case of a high Lorentz factor, the velocity is directly rescaled.

The fiducial resolution $N_r\times N_\theta \times N_\varphi$ over the simulation volume is $N_r = 32\log_{8}(r_{\rm out})$, $N_\theta = N_\varphi = 64$. We study the effect of resolution in \autoref{sec:resolution_study} where we demonstrate that the resolution used here produces the same results as at higher resolutions, consistent with previous resolution studies of MAD systems \citep{White2019,Salas2024} and of the multizone method \citep{Cho2023}.

The fifth-order WENO reconstruction \citep{Liu1994} is employed except for the ISMR-applied cells near the poles where linear reconstruction is utilized (see \autoref{sec:smr_appendix}). 1D$_W$ inversion scheme \citep{Noble2006, Mignone2007} is used for inverting the conserved variables to primitive variables. The Courant factor is set to 0.45 of the maximum stable value.

\subsection{Technical differences with previous non-spinning BH simulations}

Several modifications have been made in the code used here compared to our earlier simulations of non-spinning BH presented in \citet{Cho2023,Cho2024}. These changes are made to ensure that jet propagation across zones in appropriately captured. First, face-centered magnetic fields are used instead of cell-centered magnetic fields. With face-centered magnetic fields, it is now possible to implement ISMR (\autoref{sec:smr_appendix}) and also simplify the magnetic field prescription at boundaries (\autoref{sec:magnetic_boundary_conditions}). Second, thanks to ISMR, we now employ a coordinate system with enhanced $\theta$-resolution near the poles to better resolve the jet propagation. Third, instead of annuli with a fixed logarithmic radial width, we now use annuli with a fixed outer radius $r_{\rm{out}}$. This prevents the outgoing jet from strongly interacting with internal boundaries. Fourth, the run duration for each zone is chosen to be a fixed number of steps instead of a pre-determined characteristic timescale. This allows the simulation to naturally follow changes in characteristic timescales which can be dramatically different between V-cycles depending on the presence/absence of a relativistic jet. Applying the new simulation set-up to the problem we considered in our previous work viz., non-spinning $a_*=0$ magnetized Bondi accretion with Bondi radius $R_B\approx 2\times 10^5\,r_g$, we confirm that we reproduce the main results of \citet{Cho2023,Cho2024} to within a factor of 2. The small residual differences could arise from the change of the coordinate system or the shift from cell-centered to face-centered magnetic fields.

\subsection{Diagnostics}\label{sec:diagnostics}
We closely follow the analysis methods described in \citet{Cho2024}. For any given fluid variable $X$, its shell averaged value $\langle X\rangle$ at a given radius is defined as
\begin{equation}\label{eq:def_avg}
    \langle X \rangle (r) \equiv \frac{\iint X \rho \sqrt{-g} ~ d\theta \,d\varphi }{ \iint \rho \sqrt{-g} ~ d\theta \,d\varphi}\ ,
\end{equation}
and its time averaged version $\overline{\langle X\rangle}$ is represented by an overbar. The density average $\langle \rho \rangle$ is an exception where it is not additionally density-weighted.

The accretion rate is defined as:
\begin{equation}
    \dot{M}(r) \equiv -\iint \rho u^r\sqrt{-g}\,d\theta \,d\varphi,
\end{equation}
and the feedback efficiency is calculated as:
\begin{equation}\label{eq:eta}
    \eta(r) \equiv \frac{(\dot{M}(r) - \dot{E}(r))}{ \dot{M}(10\,r_g)}\ ,
\end{equation}
where $\dot{E}(r)\equiv \iint T^r_t\sqrt{-g}\,d\theta\,d\varphi$ is the total energy inflow rate including the rest-mass energy. The electromagnetic contribution to the feedback efficiency is 
\begin{equation}
\label{eq:etaEM}
    \eta_{\rm EM}(r) \equiv \frac{(\dot{M}(r) - \dot{E}_{\rm EM}(r))}{\dot{M}(10\,r_g)}\ ,
\end{equation}
where $\dot{E}_{\rm EM}(r) \equiv \iint T^r_{t, \rm{EM}}\sqrt{-g}\,d\theta\,d\varphi$ and $T^r_{t,\rm{EM}}\equiv b^2u^r u_t-b^r b_t$ only considers the electromagnetic terms in $T^r_t$. The dimensionless magnetic flux parameter is:
\begin{equation}
    \phi_b (r) \equiv \sqrt{\frac{4\pi}{\dot{M}(10\,r_g)\,r_g^2}}  \iint \frac{|B^r|}{2} \sqrt{-g} \, d\theta\,d\varphi.
\end{equation}
The time averaged efficiencies $\overline{\eta}, ~\overline{\eta}_{\rm EM}$ and magnetic flux $\overline{\phi_b}$ are calculated by individually time averaging $\dot{M}(10\,r_g)$ and the rest before normalizing.

Shell-averages are performed by excluding one layer of cells adjacent to each pole to avoid any numerical contamination from the polar boundary conditions. This does not apply to calculations of conserved quantities like $\dot{M}$, $\dot{E}$, $\dot{E}_{\rm EM}$, $\eta$, $\eta_{\rm EM}$, $\phi_b$, where the summation extends over all $\theta$.

Throughout, the time averaging is performed over the last 1/5 of the total runtime and is performed for each zone over the last half of each zone visit. When plotting, the time averaged radial profiles of the zones are stitched together in a similar fashion as in \citet{Cho2024}. Each zone-$i$'s profile is taken from $8^{0.5}r_{i,\rm in}$ to $8^{1.5} r_{i,\rm in}$ except for the zones at each ends (zones $0$ and $n-1$) where the profiles are calculated up to the innermost and outermost boundary, respectively. This plotting routine is not applied to small scale simulations in Section~\ref{sec:small_scale}. Instead, the innermost zone-$0$ profiles are time averaged without combining different zone averages. Since the scale separation is not large, the zone-stitched profiles and zone-$0$-only profiles are quite similar, apart from the latter being smoother.

\section{Small Scale Tests} \label{sec:small_scale}
\begin{table*}[]
\centering
 \begin{tabular}{c | c c c c c c | c} 
 \hline
Labels & Initial conditions & $n$ & $r_{\rm out}$ [$r_g$]  & Capped run duration & Runtime [$t_B$] & Coordinate System & $\eta (R_B/3)$\\ [0.5ex] 
 \hline\hline
\texttt{oz} &  \bf{B}  & - & $8^5\approx 3\times 10^4$ &  - & 50 & EKS & 0.39\\
\texttt{mz} & \bf{B} & 3 & $8^5$ & O & 50 & EKS  & 0.32\\
\texttt{oz+} & \bf{T+}  &  - & $8^5$ &  -  & 50& EKS & 1.3\\
\texttt{mz+} & \bf{T+}  &  3 & $8^5$ & O & 50  & EKS & 1.2\\ 
\texttt{mz-} & \bf{T--}  & 3 & $8^5$ & O &  50 & EKS & 0.14 \\
\texttt{mz+long} & \bf{T+}  & 4 & $8^6\approx 2\times 10^5$ & X &  450& EKS & 0.40 \\
\texttt{mz-long} & \bf{T--}  & 4 & $8^6$ & X &  450& EKS & 0.44 \\
\texttt{mz+jks} & \bf{T+}  & 3 & $8^5$ & O & 50 & JKS  & 1.2\\ [1ex]
 \hline
 \end{tabular}
 \caption{List of small-scale runs with Bondi radius $R_B\approx 400\,r_g$. The labels starting with \texttt{oz} refer to one-zone runs which are simulated using conventional methods, and the labels starting with \texttt{mz} are simulated with the multizone method. The initial conditions, {\bf B}, {\bf T+}, {\bf T--}, refer to strongly magnetized Bondi-like, weakly magnetized prograde torus-like, and weakly magnetized retrograde torus-like initial conditions, respectively. The last column lists the feedback efficiency $\eta$ in each simulation, measured at $R_B/3$.
 \label{tab:n4_run_summary}}
\end{table*}

All simulation results presented in this paper assume a BH spin of $a_*=0.9$. As in \citet{Cho2024}, we first test the new multizone set-up for spinning BHs on problems with a reduced scale separation between the BH horizon $r_H$ and the Bondi radius $R_B$. We select $R_B\approx 400 \,r_g$,\footnote{This is set by choosing a sonic radius $r_s=16\,r_g$; see Equation~A3 in \citet{Cho2024}.} which is small enough for conventional GRMHD simulations,  with no internal boundaries or zones, to be carried out. The latter are regarded as the ground-truth and referred to as the ``one-zone'' runs hereafter. In the following tests, we compare time averaged quantities between the one-zone and multizone runs. Time fluctuations are beyond the scope of this work because the statistics of variability in the multizone method is not well understood (a caveat explained in Section~\ref{sec:caveat}). We refer readers to \citet{Cho2024} for an extended discussion of the strengths and limitations of the multizone method.

Since the largest scales are hardly evolved in one-zone runs because of their large characteristic times, the multizone run duration prescription is slightly modified (only in this Section), as in \citet{Cho2024}. We reduce the number of steps taken in the outermost zone to $\approx 500\,\Delta t$ instead of $8000\,\Delta t$. This is a good approximation to the capped runtime prescription \texttt{MHDcap} used in \citet{Cho2024}.

\subsection{Effect of Initial Conditions}\label{sec:initial_condition}

\begin{figure}[ht!]
\gridline{
\fig{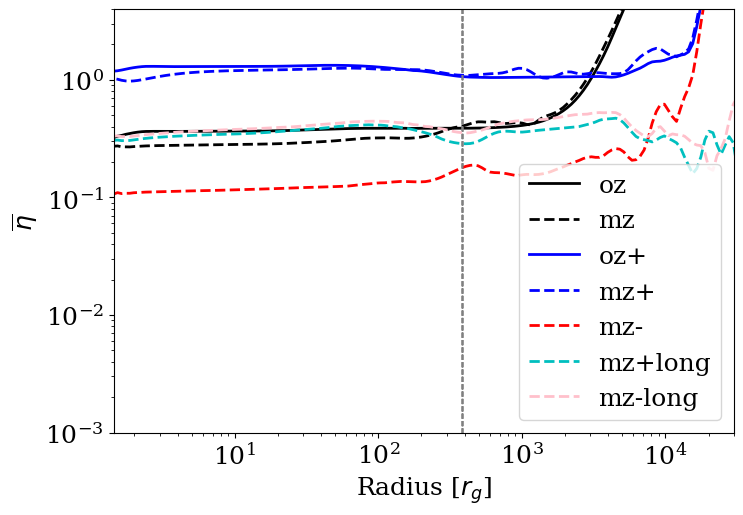}{0.48\textwidth}{}
            }   
\caption{Time-averaged radial profiles of the feedback efficiency $\overline{\eta}(r)$ in small-scale simulations (Section~\ref{sec:small_scale}) with a Bondi radius $R_B\approx 400\,r_g$. Runs with different initial conditions are labeled as in \autoref{tab:n4_run_summary}. 
Solid lines correspond to ground-truth (one-zone) runs, and dashed lines correspond to multizone runs. Black lines indicate simulations (\texttt{oz}, \texttt{mz}) initialized with the non-rotating strongly magnetized {\bf B} initial conditions (ICs), blue lines  indicate weakly magnetized prograde rotating {\bf T+} ICs (\texttt{oz+}, \texttt{mz+}), and a red line indicate weakly magnetized retrograde rotating {\bf T--} ICs (\texttt{mz-}). The cyan (\texttt{mz+long}) and pink (\texttt{mz-long}) lines show the long-term final states of the {\bf T+} and {\bf T--} runs. The vertical dashed gray line marks the location of the Bondi radius $R_B$. \label{fig:n4ICcompare}}
\end{figure}

There are three classes of ICs used in this test, henceforth referred to as {\bf B}, {\bf T+}, and {\bf T--}. The first, {\bf B}, is the spherically symmetric Bondi-like gas configuration used in \citet{Cho2023,Cho2024}. The density is initialized as $\rho_{\rm init}(r) \propto (r+R_B)/r$, which scales as $\propto r^{-1}$ at $r<R_B$ and is constant at $r>R_B$.\footnote{Even though the initial density profile resembles the final profile, \citet{Cho2023} demonstrated that the same final density profile is obtained independent of the initial density profile adopted.} The temperature $T$ and four-velocity $u^\mu$ are initialized to the general relativistic hydrodynamic analytic solution \citep{Michel1972,Shapiro1983}. Magnetic fields are initialized with a purely azimuthal vector potential $A_\varphi(r,\theta) = b_z(r+R_B)\sin{\theta}/2$,
which generates a nearly vertical field geometry. The normalization $b_z$ is chosen such that the system is strongly magnetized, with constant initial plasma-$\beta\equiv 2\rho T/b^2\sim 1$ across radii. There is zero angular momentum to begin with, though even with a non-zero initial angular momentum, the strong magnetic fields associated with $\beta\sim 1$ efficiently remove the initial angular momentum (\citealt{Cho2024,Cho2025}, see also \citealt{Chatterjee2022}).

As some studies have shown that the initial angular momentum of the gas is important for producing strong jets (\citealt{Kwan2023,Lalakos2024,Galishnikova2025}, but note \citealt{Cho2025}, who argue that the effect becomes less pronounced at late times), we additionally explore two other types of ICs, {\bf T+} and {\bf T--}, which are initialized with coherent gas rotation.
These ICs qualitatively resemble a rotating \citet{Fishbone1976} torus, the conventional IC for most GRMHD simulations in the literature. For the prograde torus-like IC, {\bf T+}, three changes are made compared to the Bondi-like IC {\bf B}: 1) the polar region is evacuated by setting $\rho_{\rm init}(r,\theta)\propto f(\theta)(r+R_B)/r$, where $f(\theta)= 1$ for $\theta \in [0.4,\pi-0.4]$  and $f(\theta) = 0.01$ elsewhere, 2) the initial magnetic field is weak, corresponding to plasma $\beta\sim 100$, and 3) there is an initial rotation with $u^\varphi = 0.5\,r^{-3/2}$ such that the gas co-rotates with the BH. The retrograde torus-like IC, {\bf T--}, is identical to {\bf T+}, except that the gas counter-rotates with the BH, $u^\varphi=-0.5\,r^{-3/2}$.

\autoref{fig:n4ICcompare} shows the results of using the three different initial configurations. Each run's set-up and the resulting efficiency $\eta$ measured at $R_B/3$ are summarized in Table~\ref{tab:n4_run_summary}. The first thing to note is that the multizone simulations (dashed lines in \autoref{fig:n4ICcompare}) reproduce almost perfectly the corresponding ground truth one-zone (solid lines) results for different ICs. The two {\bf B} IC runs (black lines, \texttt{oz} and \texttt{mz}) have feedback efficiencies $\eta\sim 0.3-0.4$, and the two {\bf T+} IC runs (blue lines, \texttt{oz+} and \texttt{mz+}) have efficiencies $\eta\sim 1.2$. The good agreement validates that the new multizone set-up used in this paper (Section~\ref{sec:numerical_method}) captures jet physics and feedback efficiency satisfactorily.

Secondly, the three different ICs, {\bf B}, {\bf T+}, and {\bf T--}, result in three distinct jet efficiencies. As {\bf T+} and {\bf T--} ICs are similar to standard GRMHD  torus simulations, we can compare the efficiencies we obtain for these ICs with previously published results for torus ICs. Our prograde torus-like {\bf T+} runs (\texttt{oz+}, \texttt{mz+}) and retrograde torus-like {\bf T--} run (\texttt{mz-}) give $\eta\sim 1.2$ and $\eta\sim 0.14$, which are close to $\eta=1.3$, 0.19 obtained for $a_*=\pm0.9$ in \citet{Narayan2022}, and $\eta=1.2$, 0.17 reported  in \citet{Cho2025}. The agreement is remarkable, considering that our torus-like ICs, {\bf T+} and {\bf T--}, only approximately resemble the \citet{Fishbone1976} torus ICs used in the previous studies.

\begin{figure}[]
\gridline{
\fig{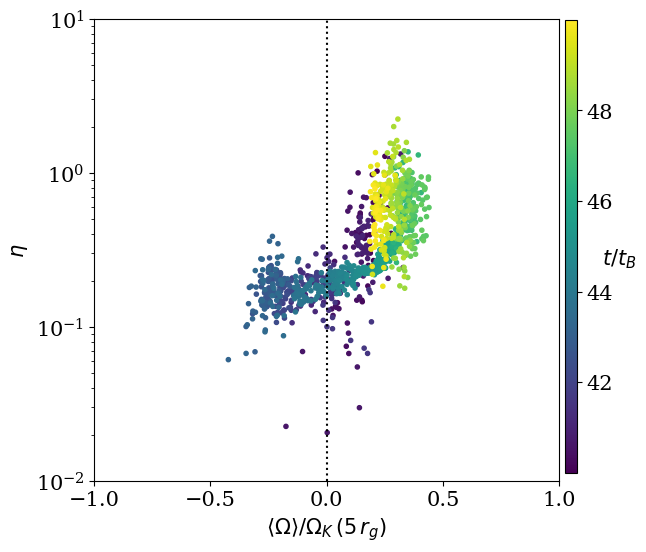}{0.48\textwidth}{}
            }   
\caption{Correlation plot of the efficiency $\eta$ against the shell-averaged angular velocity $\langle \Omega\rangle$ measured at $r=5\,r_g$. The data correspond to the strongly magnetized one-zone run with \texttt{B} ICs (model \texttt{oz}), and cover one rotation flip episode in the simulation. The color of the dots traces time $t$ divided by the Bondi time $t_B$. \label{fig:scatter_omega_eta}}
\end{figure}

Finally, we note that, when we use the {\bf B} IC, the efficiency $\eta\sim 0.3-0.4$ lies in between the values of $\eta$ for the prograde {\bf T+} and retrograde {\bf T--} ICs. Since the {\bf B} IC is strongly magnetized (initial $\beta\sim1$, compared to 100 for {\bf T+} and {\bf T--}), coherent rotation of the gas cannot survive, as demonstrated in \citet{Cho2024} and \citet{Cho2025}. As a result, the sign of the angular velocity continuously flips throughout the simulation.\footnote{We have confirmed that even with initially co- or counter-rotating gas, the angular momentum is quickly removed if we use strong magnetic field ($\beta\sim1$), so that the same final state is obtained.} If the system constantly alternates between prograde and retrograde rotation states, it is natural for the average efficiency to lie in between the two extreme. 

Figure~\ref{fig:scatter_omega_eta} shows the instantaneous efficiency $\eta$ for the strongly magnetized {\bf B} IC run (\texttt{oz}) plotted against the shell-averaged angular velocity $\langle\Omega\rangle$, both measured at $5\,r_g$. The angular velocity $\langle\Omega\rangle$ flips sign over the course of the simulation. When $\langle \Omega \rangle <0$ (counter-rotating with respect to the BH spin), the efficiency is similar to that of a retrograde torus with  $\eta\sim 0.1-0.2$. On the other hand, when  $\langle\Omega\rangle>0$ (corotating), the efficiency fluctuates over a wide range from $\eta\sim 0.1$ up to $ \sim 2$, which encompasses the efficiency $\eta\sim1.2$ seen in prograde torus runs. \autoref{fig:scatter_omega_eta} is very similar to Figure~12 in \cite{Cho2025}, which corresponds to a strongly magnetized ($\beta\sim1$) torus simulation (note that the quantities $\eta$ and $\phi_b$ in the two plots are related by $\eta\sim (\phi_b/50)^2$ for $|a_*|=0.9$ from BZ theory).

\subsection{Unique IC-independent final state}\label{sec:unique_steady_state}

Recently, \citet{Cho2025} suggested that the rotational to magnetic energy ratio in the initial state, $\mathcal{R}\equiv \rho r^2\Omega^2/b^2$, measured at the characteristic radius of the simulation (Bondi radius for the present simulations),  is an important parameter for determining the amplitude of time variability. In particular, the time when strong variability and jet intermittency first emerge, $t_{\rm var}$, appears to increase with increasing rotational dominance ($\mathcal{R}$).
Our {\bf T+} and {\bf T--} runs have $\mathcal{R}\approx20$, and according to \citet{Cho2025}'s prediction, the runs should start fluctuating strongly after $t_{\rm var} \sim 40-50\, t_B$. Models \texttt{oz+}, \texttt{mz+}, \texttt{mz-} were run for $50\,t_B$, which is marginal. On the other hand, the {\bf B} IC has no gas rotation initially, $\mathcal{R}=0$, so strong variability should set in quickly according to \citet{Cho2025}, as indeed it does. If we could run the models initialized with tori-like {\bf T+}/{\bf T--} conditions a factor of a few longer, we expect them to start fluctuating wildly and to resemble {\bf B} more closely.

The beauty of the multizone method is its ability to ``fast-forward'' a simulation to longer times without a steep increase in computational cost. We take advantage of this feature by merely adding one more zone to the torus-like {\bf T+} and {\bf T--} IC runs and studying their long-term evolution. In the same spirit, we also do not cap the number of timesteps in the largest zone. These extended runs are referred to as \texttt{mz+long} and \texttt{mz-long} respectively in \autoref{tab:n4_run_summary}. They have been run 9 times longer ($450\,t_B$) than the corresponding \texttt{mz+} and \texttt{mz-} runs  ($50\,t_B$). 

The final states of both the {\bf T+} and {\bf T--} initial states (cyan and pink lines in \autoref{fig:n4ICcompare}) in this test converge to a feedback efficiency $\eta\sim 0.4$, similar to the efficiency of the {\bf B} IC, and very different from their early time efficiencies ($\eta=1.2, ~0.14$). The jet also becomes intermittent, with strongly fluctuating $\eta$ and $\phi_b$, and the angular velocity $\Omega$ reverses direction. This is a strong validation of \citet{Cho2025}'s suggestion.

As proposed by \citet{Cho2025}, even initially rotating and weakly magnetized systems, like {\bf T+} and {\bf T--}, which have large values of $\mathcal{R}$, will eventually accumulate sufficient magnetic flux to match models like {\bf B} (which has $\mathcal{R}=0$), and will exhibit strong variability. This suggests that one strategy to efficiently reach the final state is to provide strong magnetic fields to begin with, like the {\bf B} IC's initial $\beta\sim 1$, so that a strongly fluctuating state with intermediate efficiency $\eta\sim 0.3-0.4$ is obtained within $50\,t_B$ (like models \texttt{oz}, \texttt{mz}).

We note that small scale \citet{Fishbone1976} tori have a limited initial gas and magnetic field budget, so evolving tori for very long times using the multizone method will deplete gas and magnetic field. The only way to avoid this is by starting with a  large enough torus, as in \citet{Cho2025}.
Our torus-like ICs {\bf T+} and {\bf T--}, with their Bondi-like set-up, do not face such a limitation.

\subsection{Effect of Choice of Coordinate System}\label{sec:coord_compare}

\begin{figure}[h]
\gridline{
\fig{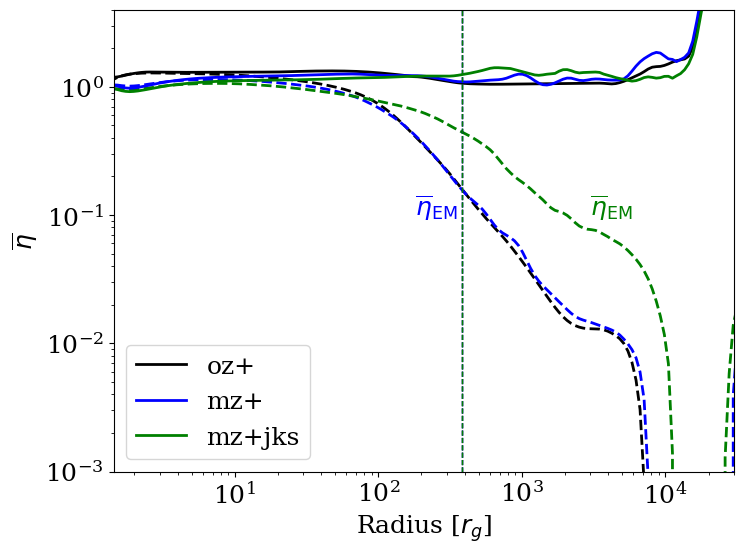}{0.48\textwidth}{}
            } 
\caption{The total feedback efficiency $\eta$ (solid lines) and the electromagnetic feedback efficiency $\eta_{\rm EM}$ (dashed lines) for different coordinate systems. The black and blue lines (\texttt{oz+} and \texttt{mz+}) correspond to the uniform polar grid EKS and the green lines (\texttt{mz+jks}) correspond to the collimating polar grid JKS. While JKS's high resolution at the poles helps preserve the Poynting flux $\eta_{\rm EM}$ to larger distances compared to EKS, the total efficiency $\eta$ is insensitive to the choice of coordinate system. \label{fig:n4coordcompare}}
\end{figure}

Because of our use of ISMR in the present simulations, we have significantly better $\theta-$resolution near the poles compared to our previous work on non-spinning BHs \citep{Cho2023,Cho2024}. Here we explore how well the new grid can capture the physics of relativistic jets. Our fiducial coordinate system is EKS, where the $\theta$ grid is uniformly spaced. However, observations indicate that jets have a quasi-parabolic morphology with jet cylindrical radius scaling as $\propto z^{0.6}$, where $z$ is the axial distance along the jet \citep[e.g.,][for the jet in M87]{Nakamura2018}. Thus the opening angle of the jet collimates as $\theta_{\rm jet}\propto r^{-0.4}$, so that a grid with a fixed $\theta$ resolution at the poles will eventually under-resolve the jet at large distances. 

This motivated us to try the new JKS coordinate system described in Appendix~\ref{sec:JKS}, which collimates the resolution at the poles at large distances to match the jet shape. We repeat the run \texttt{mz+} in JKS coordinates and refer to it as \texttt{mz+jks}. We choose the prograde {\bf T+} IC for this test because it generates the most powerful and coherent jet with $\eta\gtrsim1$.

Figure~\ref{fig:n4coordcompare} compares the feedback efficiency $\eta$ between the EKS (\texttt{oz+} in black and \texttt{mz+} in blue) and JKS (\texttt{mz+jks}, green) coordinate systems. The solid lines show the total efficiency $\eta$, where we see that the results from the three runs match each other closely up to the Bondi radius. In contrast, the dashed lines, which focus on the electromagnetic contribution to the efficiency, $\eta_{\rm EM}$ (\autoref{eq:etaEM}), show that JKS preserves the Poynting flux to larger distances compared to the two EKS runs. This suggests that, when the jet is under-resolved by the grid, the numerical scheme converts electromagnetic power to fluid energy, but without affecting the total jet power. 

We conclude that the uniform polar grid EKS is sufficient for studying the total feedback efficiency $\eta$ of Bondi-like accretion on spinning BHs. However, if we wish to study details such as the partition of jet energy between fluid and electromagnetic components, or the jet morphology at large distances, we require a collimating coordinate system such as JKS, but at the cost of needing much larger computational resources. Since our main goal is to extract the total feedback efficiency $\eta$, we use EKS as our fiducial coordinate system in the rest of this paper.

\subsection{Effect of internal magnetic boundary conditions}\label{sec:bc_comparison}
\begin{figure}[]
\gridline{
\fig{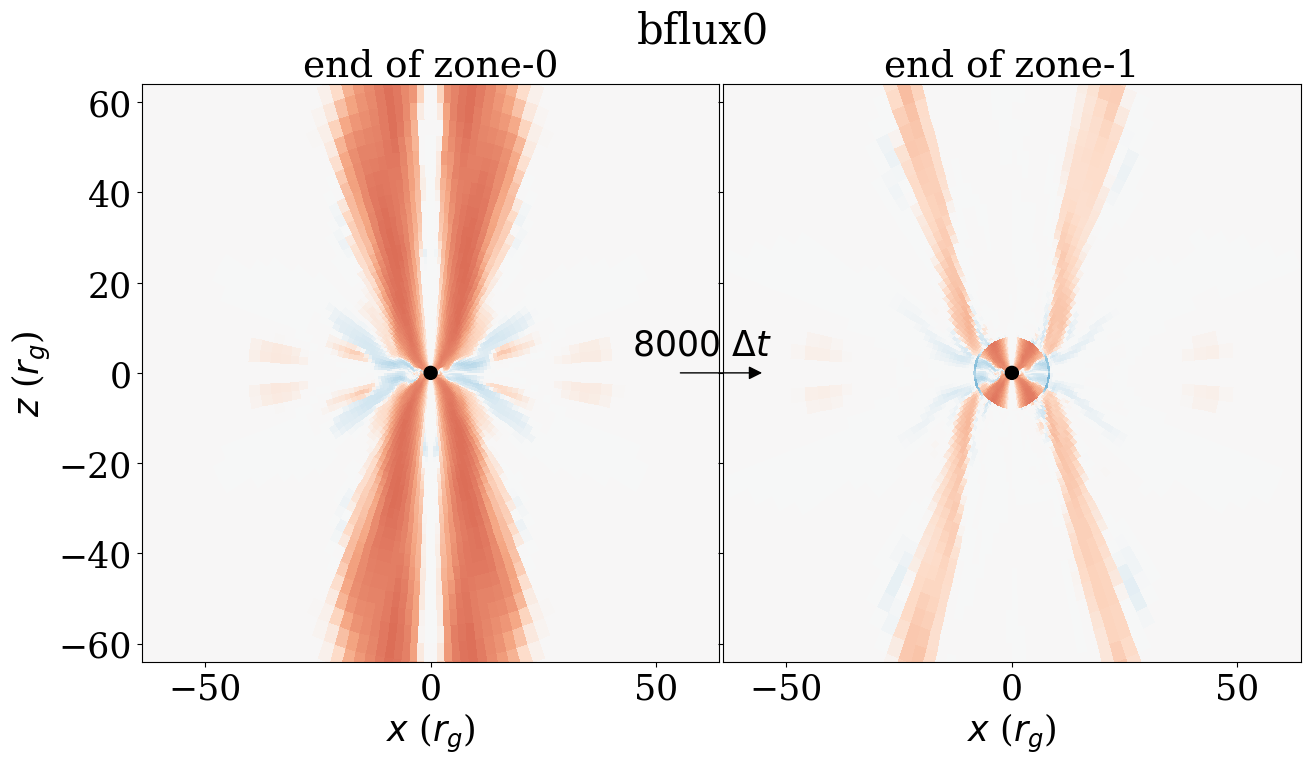}{0.48\textwidth}{}
}
\gridline{
\fig{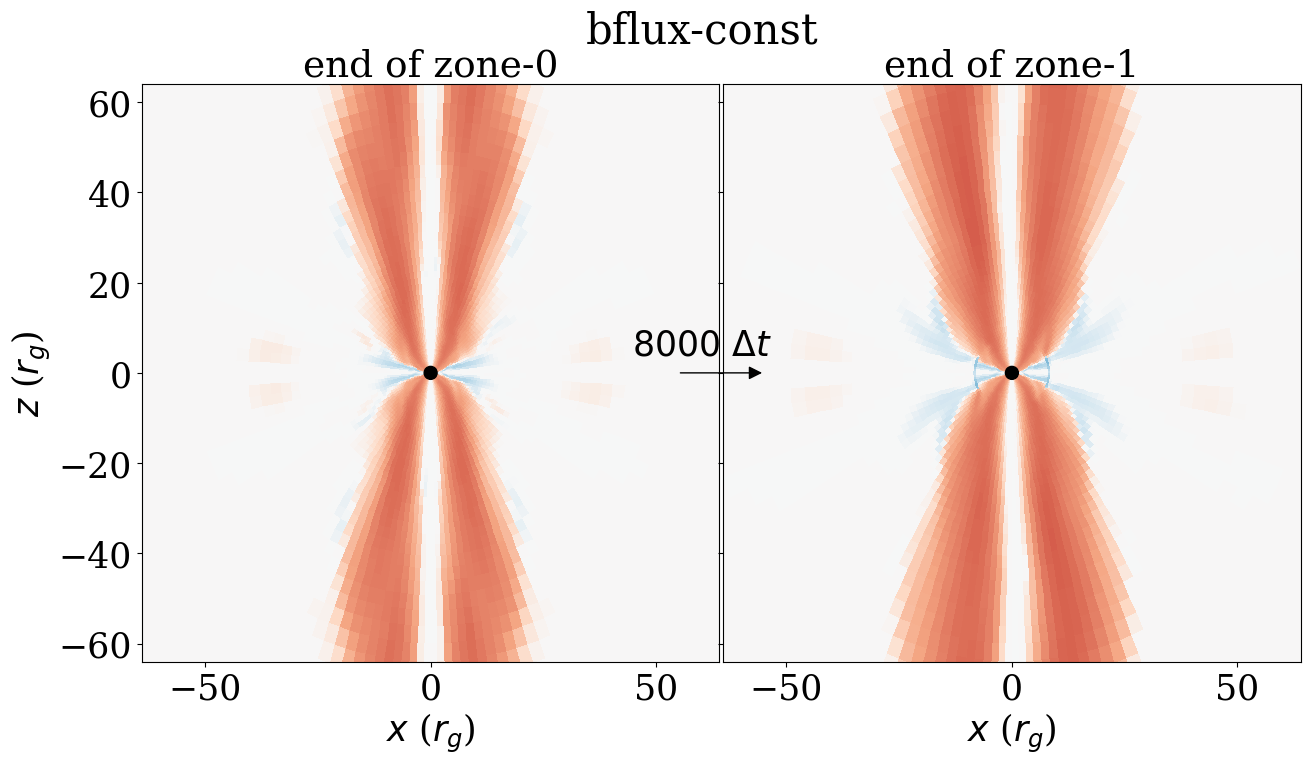}{0.48\textwidth}{}
            }   
\caption{Snapshots of the electromagnetic energy flux $T^r_{t, \rm{EM}}\sqrt{-g}$ at the end of running zone-$0$ (\emph{left}) and zone-$1$ (\emph{right}) when using the \bflux{} (\emph{top}) and \bfluxc{} (\emph{bottom}) prescriptionss. \bfluxc{} preserves the electromagnetic power of the jet across the internal boundary at $8\,r_g$, whereas \bflux{} does not.}  \label{fig:bflux_snapshot}
\end{figure}
Here we revisit the effect of the two different magnetic boundary conditions at internal boundaries, \bflux{} and \bfluxc{}, described in \autoref{sec:magnetic_boundary_conditions}. For a non-spinning BH ($a_*=0$), we found identical steady states using the two prescriptions in \citet{Cho2024}. However, there was no jet, nor any significant rotation, in those models. Since we know that the prograde {\bf T+} IC model with spin $a_*=0.9$ produces a high efficiency jet and maintains coherent rotation around the poles, we anticipate different behavior. 

Figure~\ref{fig:bflux_snapshot} shows snapshots of $T^r_{t,{\rm EM}}\sqrt{-g}$ when the \bflux{} (top row) and \bfluxc{} (bottom row) prescriptions are used with the \texttt{mz+jks} set-up. The left and right panels show the configurations at the end of running zone-$0$ and zone-$1$, respectively. With the \bflux{} prescription, when zone-1 is run, the magnetic field is anchored rigidly to the internal boundary at $r_{1,{\rm in}}=8\,r_g$. This results in extinguishing the Poynting flux beyond $r_{1,{\rm in}}$, which is obviously unphysical. On the other hand, the \bfluxc{} prescription allows an axisymmetric rotation of the magnetic field, so in this case the electromagnetic jet power continues smoothly across the internal boundary. We conclude that, for simulating accretion on spinning BHs with the multizone method, it is essential to use the \bfluxc{} prescription.

\section{Jets in Large Scale Simulations}\label{sec:large_scale}

Section~\ref{sec:small_scale} focused on test problems with an artificially small Bondi radius, $R_B\approx 400\,r_g$. Real SMBH-galaxy systems have a much larger $R_B\approx 10^5-10^6\,r_g$. Here we apply the multizone method to systems with a sequence of increasing Bondi radii, $R_B\approx 400\,r_g$, $2000\,r_g$, $2\times 10^4\,r_g$, and $2\times 10^5\,r_g$. These runs are labeled \texttt{4e2}, \texttt{2e3}, \texttt{2e4}, and \texttt{2e5}, respectively. The goal is to study the effect of the Bondi radius on both BH feeding ($\dot{M}$) and energy feedback ($\eta$).

Since small scale experiments show that the final state is reached faster with a stronger initial magnetic field (Section~\ref{sec:unique_steady_state}, also \citealt{Cho2025}), the present runs are initialized with the strongly magnetized {\bf B} IC. The runtime of the outermost zone has no cap applied in order to fully utilize the multizone method's ability to accelerate simulations. \autoref{tab:run_summary} summarizes the different set-ups for the four runs\footnote{The \texttt{4e2} run initialized with the {\bf B} IC is a fast-forwarded version of the \texttt{mz} run in \autoref{tab:n4_run_summary}. The jet efficiency is slightly smaller than \texttt{mz}, probably due to small statistical differences in the time spent in strong/weak jet phases. Also, the \texttt{4e2} run is terminated at an earlier time than the other runs to prevent strong BH feedback impacting the outermost radius $r_{\rm out}$. The other runs can evolve safely for a longer duration because $r_{\rm out}$ is much further away from $R_B$.}, along with the resulting accretion rate $\dot{M}(r_H)$ at the horizon, the feedback efficiency $\eta$ at $R_B/3$, and the jet power in units of $\dot{M}_B c^2$ obtained by combining the $\dot{M}$ and $\eta$ results. The efficiency is measured slightly inside $R_B$ because the gas dynamics changes across the Bondi radius.

\begin{table*}[]
\centering
 \begin{tabular}{c | c c c c c | c c c} 
 \hline
Labels & $R_B$ [$r_g$] & $r_s$ [$r_g$] & n & $r_{\rm out}$ [$r_g$]  &  Runtime [$t_B$] & $\bar{\dot{M}} (r_{H})$ [$\dot{M}_B$] & $\bar{\eta} (R_B/3)$ & Jet power $\dot{E}_{\rm fb}$\\[0.5ex] 
 \hline\hline
\texttt{4e2} & $\approx 400$ & 16 & 4 & $8^6\approx 2\times 10^5$ &  400 & 0.088 & 0.25 & $2.2\times 10^{-2}\,\dot{M}_B c^2$ \\
\texttt{2e3}& $\approx 2000$ & 32 & 5 & $8^9\approx 10^8$ &  700 & 0.052 & 0.29 & $1.5\times 10^{-2}\,\dot{M}_B c^2$\\
\texttt{2e4}& $\approx 2\times 10^4$ & 100 & 6 & $8^9\approx 10^8$ &  700 & 0.023 & 0.29 & $6.7\times 10^{-3}\,\dot{M}_B c^2$\\
\texttt{2e5} & $\approx 2\times 10^5$ & $10^{2.5}$ & 7 & $8^{10}\approx 10^9$ & 700 & 0.0068 & 0.24 & $1.6\times 10^{-3}\,\dot{M}_B c^2$\\ [1ex]
 \hline
 \end{tabular}
 \caption{Details of BH spin $a_*=0.9$ simulations initialized with strongly magnetized Bondi-like {\bf B} initial conditions with 4 different choices of the Bondi radius. The columns give the model name, the Bondi radius $R_B$, the sonic radius $r_s$, the number of zones $n$, the outermost radius $r_{\rm out}$, and the total runtime in units of the Bondi timescale $t_B$. The last three columns list each run's accretion rate $\dot{M}$ at the horizon $r_H$, the feedback efficiency $\eta$ measured at $R_B/3$, and the total jet power $\dot{E}_{\rm fb}$ in units of $\dot{M}_Bc^2$.
 \label{tab:run_summary}}
\end{table*}

\begin{figure*}[]
\gridline{
\fig{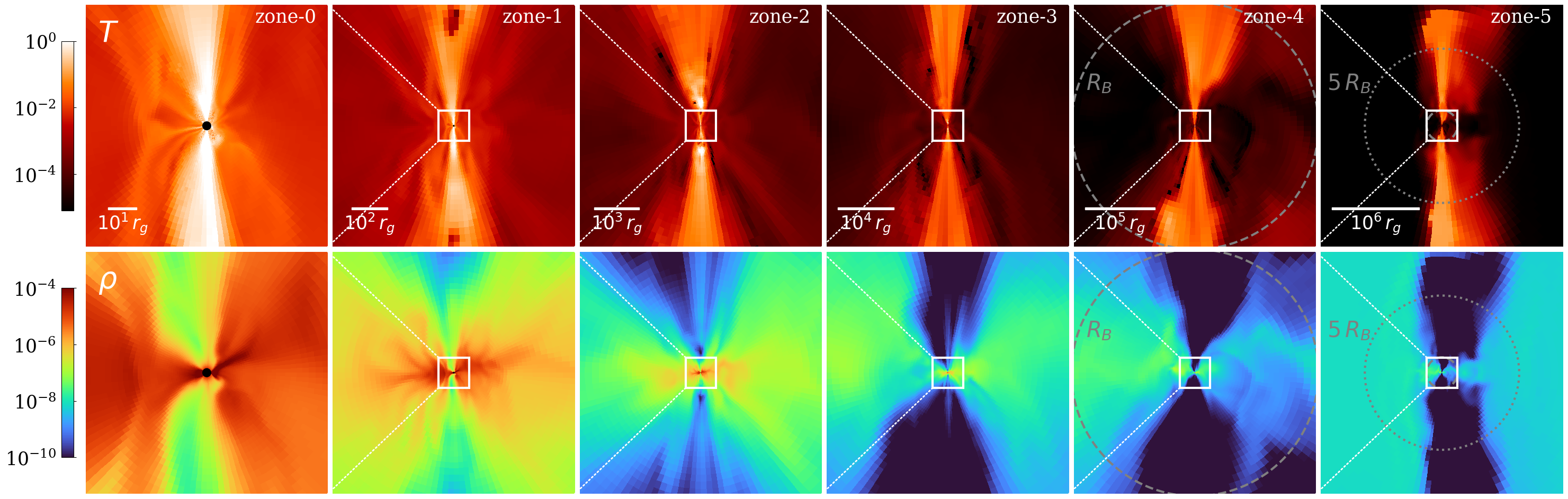}{1.\textwidth}{}
            }   
\caption{Snapshot of the BH spin $a_*=0.9$ simulation with Bondi radius $R_B\approx 2\times 10^5\,r_g$ (model \texttt{2e5} in \autoref{tab:run_summary}), selected at a time when the jet propagates beyond the Bondi radius. The upper and lower rows of panels show slices of the temperature $T$ and density $\rho$, respectively. The BH spin axis is aligned in the $+\hat{z}$ direction. From left to right, the columns go from small to large scales, focusing on the scales relevant to each zone. The relativistic jet extends from the event horizon ($\approx 0.4\,{\rm mpc}$ for M87*) to beyond $\gtrsim 5\,R_B$ ($\gtrsim 0.3\,{\rm kpc}$ for M87*), across 6 decades in radius. The scales $R_B$ and $5\,R_B$ are shown as dashed and dotted circles in the last two columns. The full movie of the simulation can be accessed at this  \href{https://youtu.be/_ecqwibOp4c}{YouTube link}.}\label{fig:snapshot}
\end{figure*}

All the simulations show strong intermittent jet activity, as expected for the late time behavior of these systems \citep{Cho2025}. Figure~\ref{fig:snapshot} shows a snapshot from the largest scale run \texttt{2e5} during a time when a powerful jet originating from the BH horizon penetrates beyond several $R_B$. The upper and lower rows correspond to the temperature $T$ and density $\rho$, respectively. Each column from left to right zooms out by a factor of $8$ to match the scale of the next larger zone. The powerful two-sided jet is aligned with the BH spin axis and is clearly visible as a large scale feature with hotter temperature and lower density.

\subsection{Effect of Varying the Bondi radius}

\begin{figure}[]
\gridline{
\fig{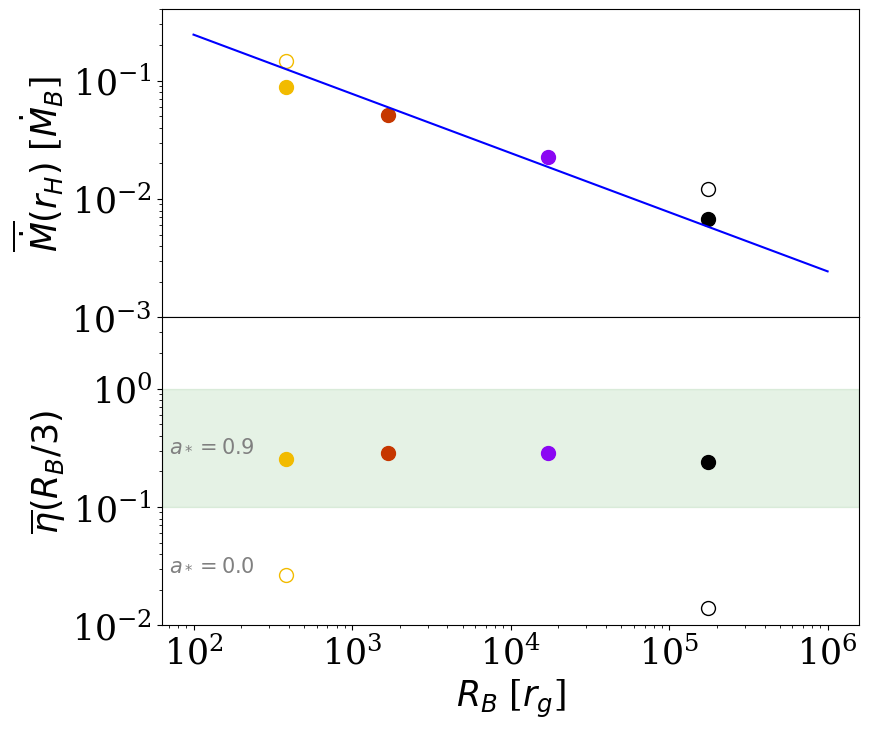}{0.45\textwidth}{}
            }   
\caption{Dependence of the time-averaged accretion rate $\overline{\dot{M}}$ in units of the Bondi rate $\dot{M}_B$ measured at the BH horizon $r_{\rm H}$, and the feedback efficiency $\overline{\eta}$ measured at $R_B/3$, as a function of the Bondi radius $R_B$ for simulations with BH spin $a_*=0.9$ (filled circles). The accretion rate scales with the Bondi radius as $\overline{\dot{M}}/\dot{M}_B\propto R_B^{-1/2}$, indicated by the blue line. The time-averaged feedback efficiency $\overline{\eta}$ is $\sim0.3$, independent of the Bondi radius $R_B$, and lies between the prograde and retrograde torus feedback efficiency values $\eta=0.1-1$, outlined by the light green background. For comparison, the spin $a_*=0$ results using the new multizone set-up are shown as unfilled circles for the smallest and largest models ($R_B\approx 400, ~2\times10^5$).}\label{fig:compare_quantity_rB}
\end{figure}

The time-averaged accretion rate $\overline{\dot{M}}(r_H)$ and feedback efficiency $\overline{\eta}(R_B/3)$ for the four models are plotted as filled circles in Figure~\ref{fig:compare_quantity_rB}. Similar to \citet{Cho2024}'s result for spin $a_*=0$, we once again find for these $a_*=0.9$ models that the accretion rate $\dot{M}$ depends on the Bondi radius as $\propto R_B^{-1/2}$. The accretion rates are approximately a factor of $\sim 2$ smaller for spinning BHs compared to the $a_*=0$ equivalents for $R_B\approx 400\,r_g$ and $\approx 2\times 10^5\,r_g$,  shown as empty circles, which we re-ran with the present code using the identical set-up as the corresponding $a_*=0.9$ runs. The moderate dependence of the accretion rate $\dot{M}$ on BH spin $a_*$ is consistent with the cyclic zoom simulations of \citet{Guo2025} and the large-scale torus simulations of \citet{Cho2025}.

Contrary to the clear dependence of the accretion rate on $R_B$, we find that the feedback efficiency $\eta$ has weak to no dependence. The feedback power we measure near the Bondi radius is $\approx 0.3 \,\dot{M}c^2$ for all four models.
Given that jet transfers energy away from the BH and the gas for accretion originates from galactic scales, it is reasonable that the jet feedback efficiency $\overline{\eta}$ should be determined by BH properties ($a_*$) and the suppression of accretion $\overline{\dot{M}}/\dot{M}_B$ should be mostly determined by galactic properties ($R_B/r_g$), although the Bondi accretion rate $\dot{M}_B$ itself depends strongly on the BH mass $M_\bullet$ as $\dot{M}_B\propto M_\bullet^2$.

This value of  $30\,\%$ efficiency lies in between those of the prograde ($\eta\sim 1$) and retrograde ($\eta\sim 0.1$) torus-like simulations discussed in Section~\ref{sec:initial_condition}.  The range of feedback efficiency between these two extremes is shown as a green background in Figure~\ref{fig:compare_quantity_rB}. We also note that a feedback efficiency $30\,\%$ is in agreement with the required minimum feedback power, $\eta\geq15\,\%$, inferred by \citet{Su2025} by comparing isolated galaxy simulations of M87* and Sgr A* with observations. 

We can combine our results for $\dot{M}$ and $\eta$ to write a single efficiency factor relating the feedback power $\dot{E}_{\rm fb}$ and the Bondi accretion rate $\dot{M}_B$:
\begin{equation}
\dot{E}_{\rm fb} = \eta_B \dot{M}_Bc^2, \quad\eta_B = \eta\,(\dot{M}/\dot{M}_B).
\end{equation}
The results from our simulations are shown in the last column of Table~\ref{tab:run_summary}. For the model \texttt{2e5}, which has a realistic  $R_B\approx 2\times 10^5\,r_g$, the feedback power is $\dot{E}_{\rm fb}\approx 2\times 10^{-3}\dot{M}_Bc^2$. 

The consistent feedback efficiency of $\eta\approx 30\,\%$ which we find across models with different Bondi radii disagrees with the recent findings of \citet{Guo2025} who report that the efficiency of the jet $\eta$ decreases with increasing $R_B$. For example, as shown in their Figure~12, the system with the largest Bondi radius has a feedback efficiency of $\eta\approx 3\,\%$, which is almost an order of magnitude weaker than what we find (their BH spin, $a_*=0.9375$, is close to our choice, 0.9). We discuss the potential causes of this disagreement in Section~\ref{sec:comparison_other_works}.

\begin{figure*}[]
\gridline{
\fig{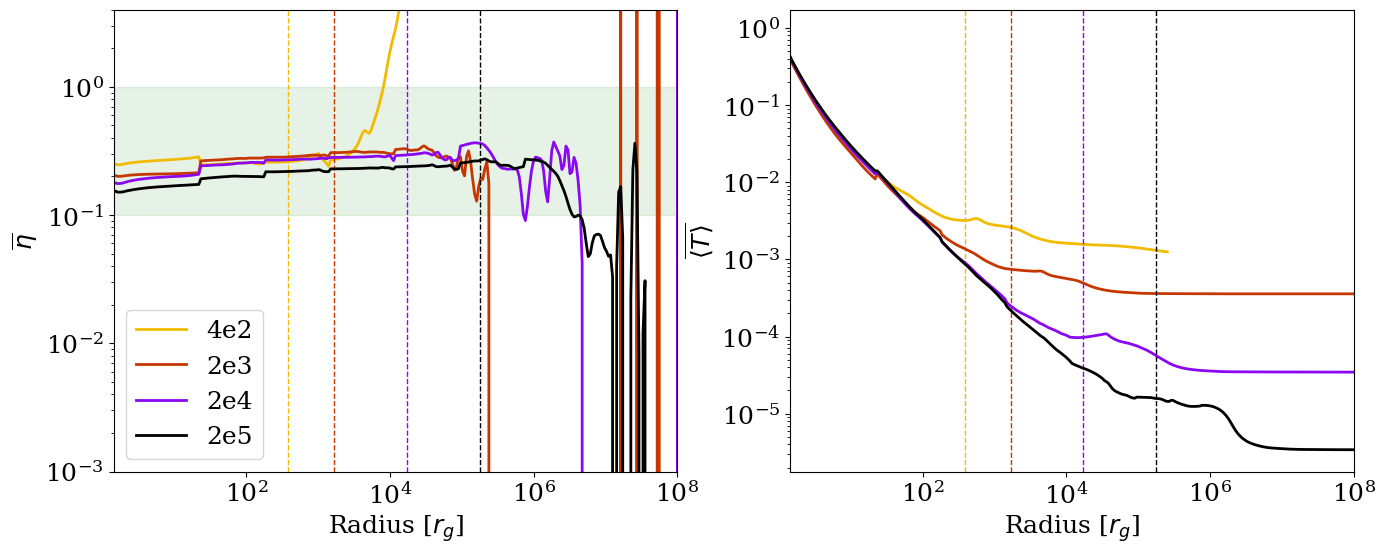}{0.8\textwidth}{}
            }   
\caption{Time-averaged radial profiles of (\emph{left}) the feedback efficiency $\overline{\eta}$, and (\emph{right}) the shell-averaged temperature $\langle \overline{T}\rangle$, for magnetized Bondi accretion with Bondi radii of $R_B\approx 400\,r_g$ (yellow), $\approx 2000\,r_g$ (red), $\approx 2\times 10^4\,r_g$ (purple), and $\approx 2\times 10^5\,r_g$ (black). The vertical dashed lines show the locations of the Bondi radii in corresponding colors. The green band displays the bounds on efficiencies from small-scale torus-like  simulations (\autoref{fig:n4ICcompare}).}\label{fig:rprofiles_a0.9}
\end{figure*}

The time averaged feedback efficiency $\overline{\eta}$ and the shell-averaged temperature $\langle \overline{T}\rangle$ profiles of our four runs are shown in Figure~\ref{fig:rprofiles_a0.9}. The feedback efficiencies $\eta$ for the four runs in the left panel all lie between the prograde and retrograde values ($\eta=0.1-1$, green background) at converged  radii $r\lesssim \,R_B$. The $\langle T\rangle$ profiles shown in the right panel closely follow the Bondi solution: $T\propto r^{-1}$ at $r<R_B$, asymptoting to a constant temperature $T_\infty= 1/(\gamma_{\rm ad}R_B)$ at large radii $r > 10\,R_B$. Near the Bondi radius $R_B$, however, there is a temperature bump due to energy deposited by feedback (see \citealt{Cho2023,Cho2024} for further discussion of this bump). 

\begin{figure*}[]
\gridline{
\fig{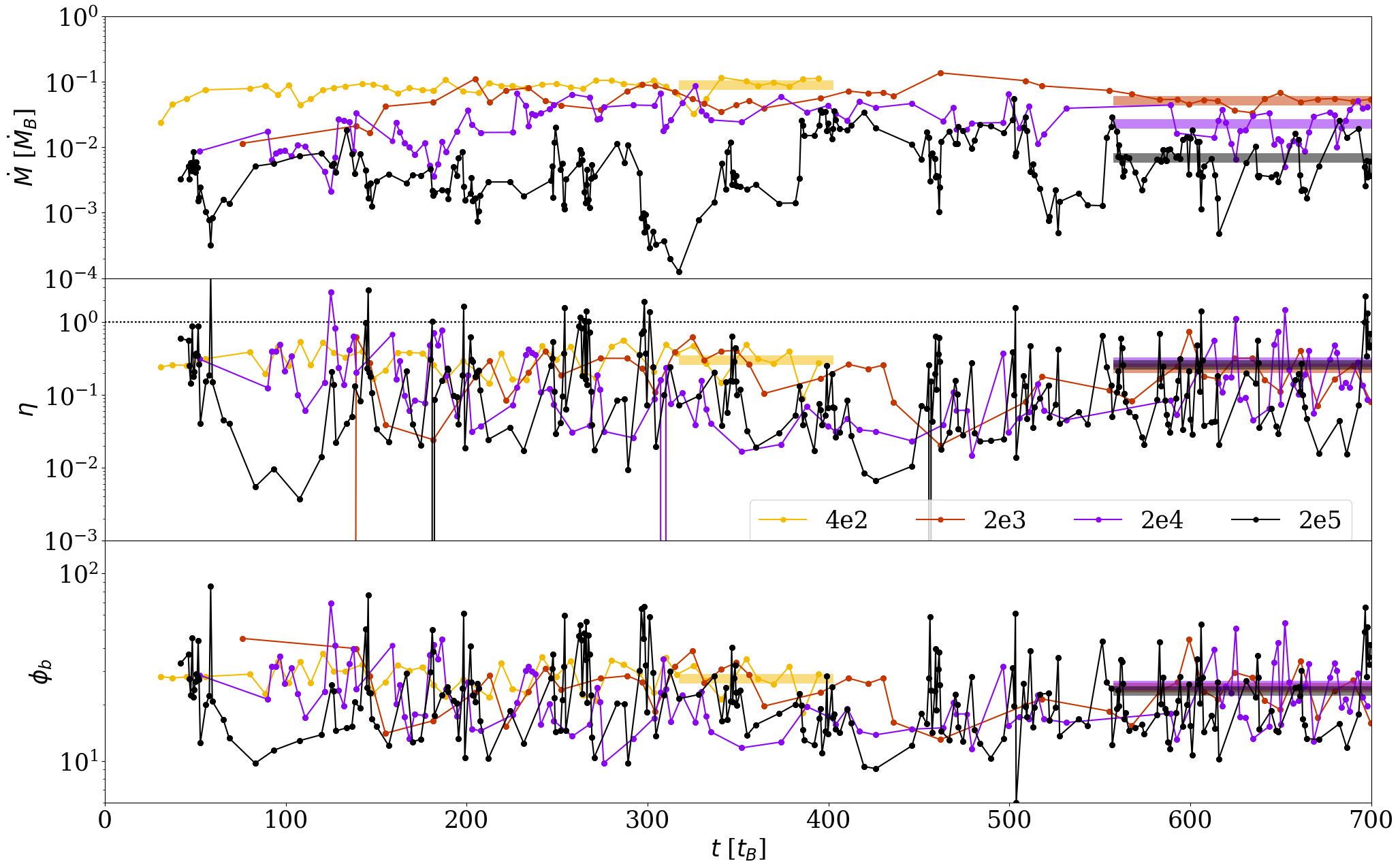}{0.95\textwidth}{}
            }   
\caption{Time evolution of the accretion rate $\dot{M}(r_H)$ in units of the Bondi accretion rate $\dot{M}_B$, the feedback efficiency $\eta(5\,r_g)$, and the dimensionless magnetic flux parameter $\phi_b(r_H)$, for four magnetized Bondi simulations with Bondi radii of $R_B\approx 400\,r_g$ (yellow curves), $2000\,r_g$ (red), $2\times 10^4\,r_g$ (purple), and $2\times 10^5\,r_g$ (black). The horizontal bars in corresponding colors indicate the time average of each quantity over the last 20\% of the simulation. While the time-averaged accretion rate $\dot{M}$ shows a decreasing trend with increasing Bondi radius $R_B$, both the feedback efficiency $\eta$ and the magnetic flux parameter $\phi_b$ remain constant across runs with different Bondi radii.}\label{fig:compare_evolution_rB}
\end{figure*}

The time evolutions of the mass accretion rate $\dot{M}(r_H)$, the efficiency $\eta(5r_g)$ and the dimensionless magnetization parameter $\phi_b(r_H)$ in the four runs (distinguished by color) are shown in Figure~\ref{fig:compare_evolution_rB}. Both $\eta$ and $\phi_b$ here are normalized by the instantaneous accretion rate $\dot{M}(10\,r_g)$ (instead of $\overline{\dot{M}}$ which we use for the time averaged $\overline{\eta}$ and $\overline{\phi_b}$ defined in Section~\ref{sec:diagnostics}).  Each dot displays an average over the last half of zone-$0$, showing one data point per V-cycle.  The time average over the last 20\% of the total simulation for each quantity is shown as a horizontal bar in corresponding colors. These plots further support our conclusion that, while the BH accretion rate $\dot{M}$ has a clear dependence on the Bondi radius ($\propto R_B^{-1/2}$), the efficiency $\eta$ as well as the strongly correlated magnetization $\phi_b$, are largely insensitive to $R_B$.

Figure~\ref{fig:compare_evolution_rB} shows that the amplitude of variability increases with increasing $R_B$, as also noted by \citet{Galishnikova2025, Guo2025}. It is possible that a system with a larger $R_B$ accommodates a larger volume of the magnetically saturated region which sources more frequent and powerful flux eruptions. The stronger perturbations may then more easily destroy the coherent structure of the jet. It is beyond the scope of the present paper to explore this further.

\subsection{Density profile along the Jet}

\begin{figure}[ht!]
\gridline{
\fig{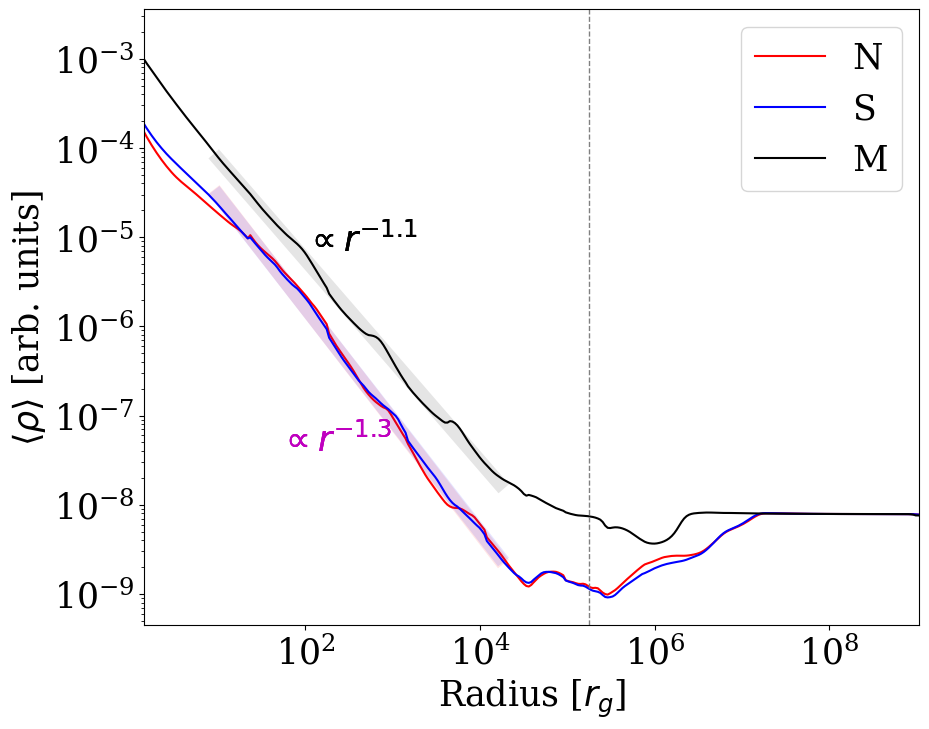}{0.5\textwidth}{}
            }   
\caption{Density profiles averaged over three ranges of the polar angle $\theta$  for the \texttt{2e5} run with Bondi radius $R_B\approx 2\times 10^5\,r_g$ (vertical gray line): north pole (N, red), $\theta=0-30^\circ$; south pole (S, blue), $\theta=150-180^\circ$; midplane (M, black), $\theta=75-105^\circ$. The jet regions (N, S) have a steeper density slope of $\rho \propto r^{-1.3}$ compared to the disk region (M) where $\rho \propto r^{-1.1}$. }\label{fig:compare_jet_disk_rho_profile}
\end{figure}

Chandra observations near the vicinity of the Bondi radius $R_B$ (M87, \citealt{Russell2018} and M84, \citealt{Bambic2023}) have revealed different radial scalings of the density parallel and perpendicular to the jet axis. Here we compare the density scaling of the jet and the disk for our largest model \texttt{2e5}. We average the density over three ranges of the polar angle, north pole (N) $\theta=0-30^\circ$, south pole (S) $\theta=150-180^\circ$, and mid-plane (M) $\theta=75-105^\circ$, and plot the time-averaged density profiles in Figure~\ref{fig:compare_jet_disk_rho_profile}. Fitting power-laws $\rho \propto r^{-\alpha}$ between $r=10\,r_g-0.1\,R_B$, we find that the jet regions N and S exhibit slopes of $\alpha=1.27\pm 0.010$ and $\alpha=1.26\pm 0.005$ and the midplane region M shows a slope of $\alpha=1.13\pm 0.006$. We obtain a similar set of slopes for the \texttt{2e4} simulation as well.

This result is similar to \citet{Bambic2023}'s observations of M84,  where they obtained a steeper density slope $\alpha\approx 1.2$ along the jet compared to $\alpha\approx 0.9$ perpendicular to the jet. However, we note that both \citet{Russell2018} and \citet{Bambic2023} measured the density scaling at $r\gtrsim R_B$, which is different from our radius range, $10\,r_g < r <0.1 R_B$. For a fair comparison, we need to extend the reliable radial range of our simulations beyond several $R_B$. This will require considering the gravitational potential of the galaxy, which can affect the density profile beyond $R_B$ \citep{Cho2024}.

The density slope of $\alpha\approx1.1$ which we find in the disk region (M) in this work is consistent with the slope $\alpha\approx 1$ found in  many BH accretion simulations in GRMHD \citep{Ressler2020b,Ressler2021,Chatterjee2022,Guo2023,Cho2023,Cho2024,Guo2024,Guo2025,Cho2025}. In particular, \citet{Chatterjee2022} measured a more accurate slope of $\alpha=1.1$ for the MAD state, which was also confirmed in \citet{Cho2025} who carefully measured the slope for a torus spanning a large range of radius. On the other hand, \citet{Xu2023}'s analytical model based on hydrodynamic turbulence predicts a shallow slope $\alpha\lesssim 0.8$. Presumably, the strong magnetic fields in the MAD state are responsible for the steeper density slope of $\alpha\approx 1.1$.

\section{Discussion}\label{sec:discussion}
\subsection{Caveats of the Multizone Method}\label{sec:caveat}
While the multizone method offers a powerful way to tackle problems with exceptionally large dynamic range, the results should be interpreted with caution. Effectively, the multizone method probes small scale evolution by only simulating a sub-sample of small scale evolution during the innermost stage (zone-$0$) in each V-cycle. Due to this sparse sampling, the method is not designed to study time variability, especially on timescales much shorter than the Bondi time. Traditional numerical methods, which capture the full statistics of the small scale evolution, are better suited for this.

We believe that the multizone method can capture the quasi-steady state or secular evolution on timescales longer than that of the largest interior radial boundary $r_{(n-1),{\rm in}}$. This is because the largest scales ($r>r_{(n-1),{\rm in}}$) always evolve continuously without being frozen in our multizone set-up. All simulations in Section~\ref{sec:large_scale} have $r_{(n-1),{\rm in}}\approx R_B$, so  we trust long-term variability on timescales $>t_B$, and time-averages over such periods. We refer readers to \citet{Cho2024} for an extended discussion of the power and limitations of the multizone method. 

\subsection{Comparison with other multi-scale simulations}\label{sec:comparison_other_works}

The vast span of scales inherent to SMBH-galaxy feedback poses a major challenge for theoretical modeling of the galactic nucleus. Prior to the deployment of the multizone method, attempts to link widely disparate scales have focused largely on tracking the accretion from galactic to BH horizon scales, thereby only capturing the inflow portion of the BH-galaxy interaction \citep{Hopkins2010,Angles-Alcazar2021,Ressler2020b,Guo2023, Hopkins2024,Hopkins2024b,Guo2024,Kaaz2025}. Tracking the other half of the communication, the back-reaction on galactic scales from the BH feedback, is tremendously expensive due to the minuscule timesteps needed to evolve BH/jet scales. Therefore, modeling the nonlinear interplay of both accretion \textit{and} feedback has only been accessible by artificially choosing a Bondi radius of $R_B\approx 100-3000\,r_g$ \citep{Lalakos2022,Kaaz2023,Lalakos2024,Galishnikova2025,Lalakos2025} as opposed to the more realistic value of $R_B\approx 10^5-10^6\,r_g$.

The multizone method which we have developed \citep{Cho2023,Cho2024} ``bridges scales'' by fully tracking the two-way communication spanning the BH-galaxy divide. The limitation imposed on traditional numerical techniques by the need to handle vastly different timescales is alleviated in this method by selectively evolving/freezing each scale. This is done in a ``V-cycle'' manner, repeated hundreds of times, to allow enough communication via accretion/feedback. The method is able to achieve steady states out to unprecedentedly large distances from the BH, with the largest dynamic range of $R_B\approx 10^7\,r_g$ achieved in \citet{Cho2024}.

Subsequently, a similar ``cyclic zoom'' method has been developed by \citet{Guo2025}. Here we briefly describe the similarities and differences between these two approaches. We also compare results from the two methods and discuss potential causes for disgreements.

Both the multizone and cyclic zoom methods significantly accelerate simulations and achieve a steady state over a wide dynamic range of scales. The core idea of the multizone method is to only intermittently evolve the small scales in the BH vicinity. This relaxes the timestep constraints and allows galactic scales to respond to any information transferred from small scales at modest numerical cost. Switching zones in a ``V-cycle'' fashion is carefully chosen to ensure ample communication between scales. A similar approach is also adopted in the cyclic zoom method. Therefore timestep constraints are similarly avoided in the two methods.

Despite this fundamental similarity, there are important differences in details of the two implementations. The most crucial difference lies in how the small scales are handled when the large scales are being evolved. The multizone method  ``freezes'' the small scales. This is easy to do with the spherical coordinate system used in this method. One numerical artifact that arises is the creation of spurious magnetic tension near the boundary between the frozen and evolving regions. However, as we have shown in Section~\ref{sec:bc_comparison}, our use of the magnetic field boundary condition \bfluxc{} prevents unsafe levels of magnetic tension and, importantly, preserves jet power across internal boundaries (it also preserves $\nabla\cdot B=0$, \citealt{Cho2024}). Furthermore, tests show that the radial profiles of quantities of interest in multizone simulations match those of the corresponding one-zone simulations (see Figure~\ref{fig:n4ICcompare}), further validating the method.

The cyclic zoom method, which uses a Cartesian grid, loosens the timestep constraint by de-refining the central grid to low resolution in each V-cycle. In doing so, small-scale structures are smoothed away, which could be troublesome when accumulating magnetic flux on the BH horizon in the MAD state. The cyclic zoom method thus requires two additional approximations in order to compensate for the lost small-scale information. First, the method freezes the de-refined fluid variables ($\rho$, $T$, $u^\mu$) to preserve the high-resolution information to some degree. But the de-refined magnetic fields continue to evolve. This violates the flux freezing condition of ideal magnetohydrodynamics, the consequences of which should to be studied further. Second, a subgrid prescription is applied to boost the electromagnetic power on small scales, to make up for the lost small-scale magnetic energy during de-refinement. 
As with our multizone approach, \citet{Guo2025} demonstrate excellent agreement between the ground truth (which we call one-zone) and cyclic zoom results for smaller scale problems.

Although both methods pass small-$R_B$ tests, when it comes to large-scale problems, $R_B$ approaching realistic values $\gtrsim 10^5r_g$, there is no ground truth, so the only option we have is to cross-compare results from the two methods. \citet{Guo2025} successfully reproduce the multizone results for non-spinning BHs reported in \citet{Cho2023,Cho2024}. This is a very welcome validation of both methods, at least for non-spinning BHs.

However, the two methods disagree to some extent when it comes to large-scale simulations ($R_B>10^5\,r_g$) with spinning BHs and jets. Large scale simulations initialized with strong magnetic fields generate a chaotic jet (as we have shown in Section~\ref{sec:large_scale}), where both methods' prescriptions are untested. Cyclic zoom's recurrent de-refinement of magnetic fields over up to 19 de-refinement levels  may potentially have a cumulative effect, which might explain the decreasing efficiency $\eta$ at each de-refinement level in Figure 11 of \citet{Guo2025}. 

\citet{Guo2025}'s low feedback efficiency of $\eta\sim 3\,\%$ at around the Bondi radius $R_B$ for BH spin $a_*=0.9375$ is an order of magnitude smaller than our feedback efficiency $\eta\sim 30\,\%$ for $a_*=0.9$.\footnote{We note that \citet{Guo2025}'s reported efficiency of $\eta\sim 10\,\%$ is measured at the horizon $r_H$. The efficiency $\eta\sim3\,\%$ we quote here is measured at $r>10^3\,r_g$, where their efficiency profile $\eta(r)$ flattens out (see Figure 11 in \citealt{Guo2025}). The value $\eta\sim 3\,\%$ is thus the actual feedback energy transferred to galactic scales. This same value is plotted in Figure 12 of \citet{Guo2025}.} Furthermore, their value of $3\,\%$ is almost comparable to our efficiency for non-spinning BHs ($\sim$ a few percent). When combined with their suppressed accretion of $\dot{M}\approx 4\times 10^{-3}\dot{M}_B$, the resulting feedback power transferred to galactic scales is only $\dot{E}_{\rm fb}\approx 10^{-4}\dot{M}_Bc^2$ (compared to our $\dot{E}_{\rm fb}\approx2\times 10^{-3}\,\dot{M}_Bc^2$ for \texttt{2e5} model). This low value is likely insufficient to supply the feedback power required in galactic/cosmological simulations to modulate star formation and BH growth  (e.g. \citealt{Su2025} require feedback efficiency of $\eta\gtrsim 15\,\%$ in isolated galaxy simulations to explain current observations of M87* and Sgr A*). 

At present, the multizone and cyclic zoom methods are the only tools available to fully bridge realistic BH-galaxy scales and track bidirectional communication. The two methods have their strengths and weaknesses, so choosing the appropriate tool for each problem and physics question of interest is important. The strength of the multizone method is the full preservation the small-scale information, while its weakness is the creation of spurious magnetic tension at internal boundaries, although tests indicate that the latter effect is not very damaging. The cyclic zoom's Cartesian grid meanwhile provides flexibility in focusing the resolution at multiple locations, but  smoothing out smaller-scale structure may bias results, especially for problems with a large dynamic range. In our view, the multizone method is suitable for problems with extremely large scale separation with a clear center, e.g., a single accreting SMBH, while cyclic zoom is preferable for simultaneously tracking multiple locations with high resolution. 

Because the ground truth for a large dynamic range problem is and will continue to be out of the reach of traditional methods, having more than one method and cross-validating them is the only way to make progress in the short term. An in-depth investigation and comparison of the multizone and cyclic zoom methods will greatly advance our understanding of BH-galaxy steady states.

\subsection{Feedback Subgrid Models for Cosmological Simulations}

Similar to the accretion subgrid formula presented in \citet{Cho2024} for non-spinning BHs ($a_*=0$), here we provide a feedback prescription that is straightforward to implement in galaxy-scale and cosmological simulations.
Putting together the results from Section~\ref{sec:large_scale} (i.e. the Bondi radius-dependent suppression of the accretion rate $\dot{M}/\dot{M}_B\propto R_B^{-1/2}$ and the fixed efficiency $\eta\approx 30\,\%$ for BH spin $a_*=0.9$), the feedback power, expressed in terms of the Bondi accretion rate $\dot{M}_B$, is
\begin{equation}\label{eq:subgrid_suggestion}
    \dot{E}_{\rm fb}\approx\begin{cases}
         2 \times 10^{-3}\left(\frac{R_B}{2\times 10^5\,r_g}\right)^{-1/2}\dot{M}_B c^2, &(a_*=0.9)\\
         2\times 10^{-4}\left(\frac{R_B}{2\times 10^5\,r_g}\right)^{-1/2}\dot{M}_Bc^2,& (a_*=0).
    \end{cases}
\end{equation}
The second formula is derived from the estimates of \citet{Cho2024} of the efficiency $\eta\approx 0.02$ and accretion rate $\dot{M}/\dot{M}_B$ for non-spinning ($a_*=0$) BHs. Note that feedback from a rapidly spinning BH ($a_*=0.9$) is predominantly powered by the jet, whereas for a non-spinning BH ($a_*=0$) the feedback is powered by reconnection near the BH \citep{Cho2023,Cho2024}. \autoref{eq:subgrid_suggestion} can be directly used in cosmological simulations where SMBH accretion is poorly resolved. Although realistic Bondi radii vary over only a modest range, $R_B\approx10^5-10^6\,r_g$, the above scaling with the Bondi radius may be especially useful to include when BHs are embedded in unusually hot or cold environments.

\citet{Su2025} found that $\dot{E}_{\rm fb} \gtrsim 1.5\times 10^{-3}\dot{M}_B c^2$ is required to match the observed accretion rate and the star formation rate in M87*. An even higher feedback efficiency appears to be needed to reproduce the observations of Sgr A*. In view of \autoref{eq:subgrid_suggestion}, the BHs in these systems might be spinning fairly rapidly.

Although \autoref{eq:subgrid_suggestion} applies only for BH spins $a_*=0,~0.9$, provided that the average feedback efficiency lies between prograde and retrograde values and is independent of $R_B$, standard small-scale torus simulations can be used to constrain the average efficiency $\eta$ for other BH spins. 
Unlike the wide range of $\eta\approx 0.1-1$ between retrograde and prograde models for $a_*=0.9$, lower BH spins have a tighter range (e.g., $\eta=0.10-0.34$ for $a_*=0.5$; \citealt{Tchekhovskoy2012,Narayan2022}).
Also, for a similar magnetic flux on the horizon $\phi_b(r_H)$, the coefficient in \autoref{eq:subgrid_suggestion} is expected to scale roughly $\propto a_*^2$ according to the BZ model. This would suggest a coefficient $\sim6\times 10^{-4}$ for $a_*=0.5$ and $\sim10^{-3}$ for $a_*=0.7$. Simulations using the multizone method for several different values of $a_*$ will enable us to improve these rough estimates and develop a generalized formula for the feedback power $\dot{E}_{\rm fb}(R_B,a_*)$. This is left for future work.

Finally, we compare the feedback power we propose in \autoref{eq:subgrid_suggestion}, which is derived from our first-principles simulations and has no adjustable parameters, to a collection of subgrid prescriptions that have been adopted, mostly by trial and error, in large-scale cosmological simulations. The models vary wildly across different studies, but restricting our comparison to Bondi-related prescriptions for the low-Eddington mode of accretion (hot accretion), the feedback power assumed in a representative sample of approaches is as follows: $\lesssim 0.2\dot{M}_B c^2$ in IllustrisTNG \citep{Weinberger2018}; $\lesssim 0.015\dot{M}_B c^2$ in EAGLE \citep{Schaye2015}; $\sim 5\times 10^{-3}\dot{M}_Bc^2$ in ASTRID \citep{Ni2022}; $\sim 3 \times 10^{-3}\dot{M}_B c^2$ in SIMBA \citep{Dave2019}; and $\sim 2\times 10^{-3}\dot{M}_B c^2$ in ROMULUS \citep{Tremmel2017}. We note that, beyond the feedback efficiencies, the detailed implementations of feedback prescription differ significantly across models and effective efficiency may be constrained by different physical or numerical conditions in each case.

While ASTRID, SIMBA, and ROMULUS utilize a level of feedback power comparable to what we predict based on our multizone simulations with BH spin $a_*=0.9$, the prescriptions implemented in IllustrisTNG and EAGLE require one to two orders of magnitude stronger feedback power.

\section{Summary and Conclusion}\label{sec:conclude}

The multizone method \citep{Cho2023,Cho2024} has opened up the possibility of directly simulating, from first principles, accretion onto a SMBH and modeling its influence on galactic scales. This is an important step forward in understanding the coevolution of BH-galaxy system. Our previous multizone work \citep{Cho2023,Cho2024} was limited to non-spinning BHs, which provide only weak feedback. In this work, we applied the multizone method to BHs with spin $a_*=0.9$ to study the BH-galaxy final state in the presence of strong feedback from relativistic jets. To enable this project, we made several modifications to the numerical method, as outlined in Section~\ref{sec:numerical_method}.

In Section~\ref{sec:small_scale}, we tested the multizone method by artificially reducing the scale separation between the BH and the galaxy by setting the  Bondi radius to $R_B\approx 400\,r_g$. This reduced scale separation makes the problem manageable even with the standard GRMHD method, which we refer to as the one-zone run, and whose results we regard as the ground truth. Time-averaged steady states for the BH spin $a_*=0.9$ problem show excellent agreement between the multizone and one-zone runs, demonstrating the validity of the multizone method even for highly spinning BH spacetimes. 

We find an interesting dependence of the results on the initial gas properties. Simulations initialized with prograde {\bf T+} and retrograde {\bf T--} torus-like conditions, where gas (i) is evacuated at the poles, (ii) is weakly magnetized, and (iii) rotates, exhibit jet feedback efficiencies of $\eta\sim 100\,\%$ and $\eta \sim 10\,\%$ respectively, after running for a duration of $50t_B$. These efficiencies are consistent with the results of conventional simulations of prograde and retrograde tori \citep[e.g., ][]{Tchekhovskoy2012,Narayan2022,Cho2025}. However, when initialized with a strongly magnetized non-rotating Bondi-like flow {\bf B}, we obtain an intermediate efficiency $\eta \sim 30\,\%$. As dynamically dominant magnetic fields efficiently transport angular momentum in the accreting gas \citep{Chatterjee2022}, the accretion flow in {\bf B} alternates between prograde and retrograde states, leading to an intermediate feedback power.

We have been able to confirm the suggestion of \citet{Cho2025} that the three different initial gas conditions, {\bf T+}, {\bf T-}, {\bf B}, should eventually converge to the same final steady state if they are evolved over a sufficiently long period of time. For this test, we accelerated the  {\bf T+} and  {\bf T-} simulations by taking advantage of the multizone method. The result is that {\bf T+} and {\bf T --}, which are weakly magnetized to begin with, finally accumulate a sufficient magnetic flux to resemble {\bf B}, and their feedback efficiencies then settle to $\eta\sim40\,\%$, similar to $\eta\sim30\,\%$ which we find (much earlier) for {\bf B}. In practice, this suggests that it might be better in the future to initialize simulations with strong magnetic fields so that they can reach their final steady states faster.

After validating the multizone method with the above small-scale tests ($R_B\approx 400\,r_g$) with spinning ($a_*=0.9$) BHs, in Section~\ref{sec:large_scale} we applied the method to highly magnetized Bondi accrretion ({\bf B} ICs)  with different Bondi radii: $R_B\approx 400,~2000,~2\times 10^4,$ and $2\times 10^5 \,r_g$. In the simulation with a realistic Bondi radius, $R_B\approx 2\times 10^5\,r_g$, we demonstrated that a relativistic jet can extend beyond several Bondi radii ($\gtrsim5\,R_B$, or $\gtrsim 0.3\,{\rm kpc}$ for M87*), depositing feedback energy on galactic scales. When comparing simulations with different choices of $R_B$, we find a dependence of the time-averaged accretion rate $\overline{\dot{M}}$ on the Bondi radius $R_B$ with a scaling $\overline{\dot{M}}/\dot{M}_B\propto R_B^{-1/2}$, consistent with the non-spinning BH results in \citet{Cho2024}. However, the time-averaged feedback efficiency $\overline{\eta}$ shows little to no dependence on the Bondi radius $R_B$; we obtain $\eta\sim30\,\%$ for all four Bondi radii. Combining these results, we predict that the jet feedback power for $a_*=0.9$ is $\dot{E}_{\rm fb}\approx 2\times 10^{-3}\,\dot{M}_B c^2$ for a realistic galaxy with $R_B\approx 2\times 10^5\,r_g$ (\autoref{eq:subgrid_suggestion}). Our estimate of the feedback power, obtained from these first-principles simulations, is comparable to the feedback power assumed in some (but not all) cosmological simulations, where the power is calibrated to match observations. 

Furthermore, we find that the amplitude of time variability in the feedback power increases with increasing $R_B$. In addition, the time-averaged density scaling is different between the jet and disk regions, with $\rho\propto r^{-1.3}$ and $\rho\propto r^{-1.1}$, respectively.

Our key results are the following.
\begin{itemize}
    \item We have shown that simulations starting with different initial conditions in the magnetic field strength (initial $\beta\sim 1$ vs 100) or the amount of rotation (0.5 Keplerian vs no rotation) eventually reach the same final state when the simulation is run long enough for sufficient magnetic fields to accumulate near the BH.
    \item The final time-averaged jet feedback efficiency $\overline{\eta}$ is determined by the properties of the BH, specifically the BH spin $a_*$, and is insensitive to where the Bondi radius $R_B$ is located. For spin $a_*=0.9$ considered in this paper, the jet feedback efficiency is $\overline{\eta} \sim 30\,\%$. Although the time-averaged efficiency $\overline{\eta}$ is independent of $R_B$, the instantaneous jet efficiency $\eta(t)$ fluctuates with increasing amplitude as $R_B$ increases.
    \item The BH mass accretion rate, on the other hand, depends on $R_B$ as $\overline{\dot{M}}/\dot{M}_B\propto R_B^{-1/2}$, but depends only modestly on the BH spin; spinning BHs accrete a factor of 2 less compared to their non-spinning counterparts.
    \item Combining the previous two findings, we provide a formula for the feedback power (\autoref{eq:subgrid_suggestion}) which can be used directly as a subgrid prescription in cosmological simulations. For a realistic Bondi radius $R_B\approx 2\times 10^5\,r_g$, and assuming a BH spin $a_*=0.9$, the feedback power is $\approx 2\times 10^{-3}\,\dot{M}_Bc^2$. The power is an order of magnitude less for $a_*=0$. Large-scale galaxy simulations need a similar or even larger feedback power to that estimated for our $a_*=0.9$ run to match observations. If we ignore the considerable systematic effects behind such a comparison, it suggests that SMBHs in the Universe are more likely to be spinning rapidly ($a_*$ close to 0.9) than slowly ($a_*$ close to 0).
    \item  The unique strength of the multizone method presented here is its ability to capture small-scale dynamics near the BH and follow the impact of the released feedback energy over a wide range of radius, all within a single simulation. This is difficult to accomplish using standard numerical techniques due to the overwhelmingly large range of timescales.
\end{itemize}

By bridging the BH and galaxy scales, the multizone method can resolve the large uncertainty currently present in the modeling of BH accretion and feedback in galaxy/cosmological simulations. Since large-scale simulations cannot resolve scales inside $R_B$, they use empirical subgrid prescriptions that are adjusted by fitting to observations. An alternative approach is to use the results from multizone GRMHD simulations and incorporate them directly as a first-principles subgrid prescription, as done in \citet{Su2025}. Furthermore, it will even be possible to directly patch GRMHD and galaxy simulations in the near future, removing the necessity of subgrid prescriptions altogether. Such direct patching was attempted in \citet{Cho2024} for non-spinning BHs by initializing GRMHD simulations with the large-scale environment from GIZMO simulations \citep[also see][]{Kaaz2025}. In a similar manner, multizone simulation data can be remapped and incorporated into galaxy simulations. The exchange of information between two simulations on very different scales, one on the galaxy scale and the other on the BH scale, can be repeated iteratively where they overlap, forming even larger overarching V-cycles. Such extensions of the multizone approach are longer-term goals.

\section*{Acknowledgements}
We thank Minghao Guo, James Stone, Eliot Quataert, Lars Hernquist, John Raymond, and Aneta Siemiginowska for valuable discussions and suggestions.
H.C., R.N., K.S., and P.N. were partially supported by the black hole Initiative at Harvard University, which is funded by the Gordon and Betty Moore Foundation (Grant \#8273.01). It was also made possible through the support of a grant from the John Templeton Foundation (Grant \#62286). The opinions expressed in this publication are those of the authors and do not necessarily reflect the views of these Foundations.
B.P. was supported by the US Department of Energy through the Los Alamos National Laboratory. Los Alamos National Laboratory is operated by Triad National Security, LLC, for the National Nuclear Security Administration of U.S. Department of Energy (Contract No. 89233218CNA000001). This work has been assigned a document release number LA-UR-XX-YYYY. This work used Delta at the University of Illinois at Urbana Champaign through allocation AST080028 from the Advanced Cyberinfrastructure Coordination Ecosystem: Services \& Support (ACCESS) program, which is supported by U.S. National Science Foundation grants \#2138259, \#2138286, \#2138307, \#2137603, and \#2138296.

\clearpage
\newpage

\appendix

\section{Magnetic boundary condition for Face-centered Magnetic Fields}\label{sec:magnetic_boundary_conditions}
The magnetic field conditions at Dirichlet radial boundaries for the cell-centered magnetic fields were described in \citet{Cho2024}. Here we describe the equivalent conditions used in this paper, where they are simpler for the face-centered magnetic fields. In the ghost cells in the radial direction including the boundary itself, the fluid variables ($\rho, u, u^\mu$) and the magnetic fields are frozen to each active zone's initial values. In contrast to the fluid variables, the magnetic fields need special prescription to additionally satisfy the no monopole  $\nabla\cdot B =0$ condition. Similarly to \citet{Cho2024}, the prescriptions are described in 2D first and then generalized to 3D. The superscript $n$ is the timestamp and subscripts indicate the spatial location in the grid with integer subscripts $i,j,k\in\mathbb{Z}$ corresponding to cell centers.

\begin{figure}[ht!]
\gridline{
\fig{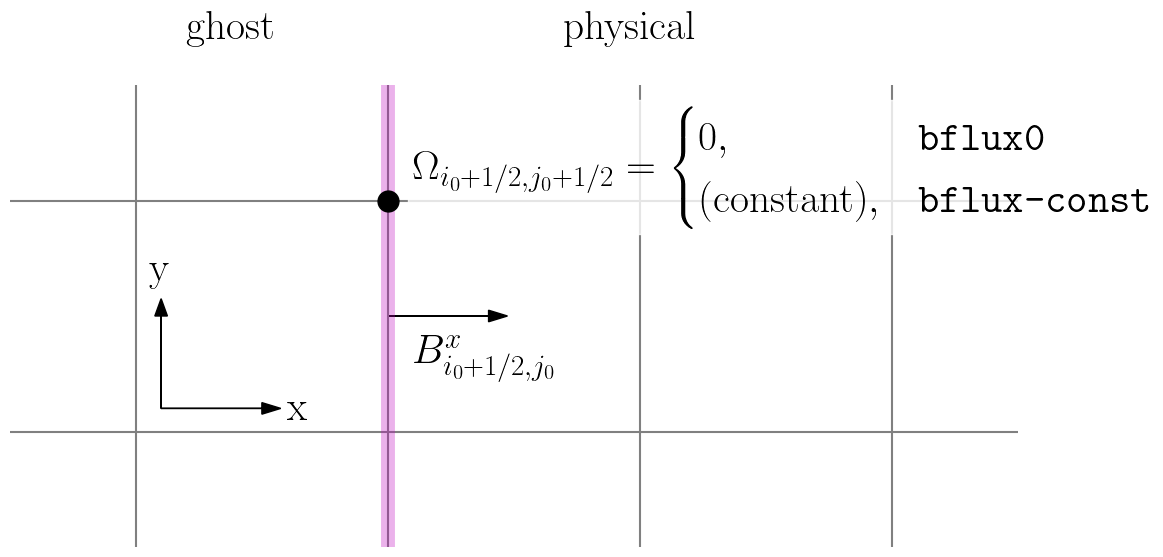}{0.6\textwidth}{}
            }   
\caption{A 2D schematic of the magnetic boundary conditions (\bflux{} and \bfluxc{}) similar to \citet{Cho2024} but for face-centered magnetic fields.
}\label{fig:bflux_schematic}
\end{figure}

In 2D, the face-centered magnetic fields are evolved following the induction equation \citep{Balsara1999}
\begin{equation*}
\begin{split}
    B^{x,n+1}_{i+1/2,j} &= B^{x,n}_{i+1/2,j} -\Delta t\frac{\Omega_{i+1/2,j+1/2} - \Omega_{i+1/2,j-1/2}}{\Delta y}, \\
    B^{y,n+1}_{i,j+1/2} &= B^{y,n}_{i,j+1/2} +\Delta t\frac{\Omega_{i+1/2,j+1/2} - \Omega_{i-1/2,j+1/2}}{\Delta x},
\end{split}
\end{equation*}
where $\Omega$ is the z-component of the electric field $E^z$ (EMF) at the corners.
The divergence of the magnetic fields is
\begin{equation*}
\begin{split}
    (\nabla\cdot B)_{i,j} &=\frac{B^{x}_{i+1/2,j} - B^{x}_{i-1/2,j}}{\Delta x} + \frac{B^{y}_{i,j+1/2}-B^{y}_{i,j-1/2}}{\Delta y}.
\end{split}
\end{equation*}
If the Dirichlet boundary condition enforces that the magnetic fields are kept constant at the $i_0+1/2$ surface (i.e. $B^{x,n+1}_{i_0 +1/2,j} = B^{x,n}_{i_0+1/2,j}$ for $\forall j$), then the magnetic divergence at the first physical cell center neighboring the Dirichlet boundary at $i_0+1,j$ evolves as
\begin{equation*}
\begin{split}
    (\nabla\cdot B)^{n+1}_{i_0+1,j} &= (\nabla\cdot B)^{n}_{i_0+1,j} + \frac{\Delta t}{\Delta x \Delta y}\left(-\Omega_{i_0+1/2,j+1/2} + \Omega_{i_0+1/2,j-1/2}\right).
\end{split}
\end{equation*}

Two different approaches can be taken to cancel out the last two terms and preserve the divergence for Dirichlet boundary conditions $(\nabla\cdot B)^{n+1}_{i_0+1,j} = (\nabla\cdot B)^{n}_{i_0+1,j}$. The \bflux{} prescription sets the EMFs at the $i_0+1/2$ boundary to zeroes and the \bfluxc{} prescription sets them to constants where the constant is the EMF averaged along the $y$ direction $\langle\Omega\rangle_y\equiv \frac{1}{n_2+1}\Sigma_{j'} \Omega_{i_0+1/2,j'+1/2}$.
\begin{align}
\Omega_{i_0+1/2,j+1/2}\text{ (2D)} =
    \begin{cases}
     0 ~ & (\bflux{})\\
     \langle\Omega\rangle_y ~ \rm{(constant)} ~ & (\bfluxc{})
\end{cases}
(\forall j)
\end{align}

Extending to 3D, \bflux{} is easily generalized as $\Omega^y_{i_0+1/2,j,k+1/2} = \Omega^z_{i_0+1/2,j+1/2,k}=0$ for $\forall j,k$. In contrast, \bfluxc{} prescription needs extra caution because of potential conflict with reflecting polar boundary conditions at $x$, $y$ corners. Since the polar (corresponding to $y-$direction) boundary condition puts EMFs to zeros $\Omega^x=\Omega^z=0$ at the boundary, $z-$component of the EMF is put to zero to avoid the conflict $\Omega^z_{i_0+1/2,j+1/2,k}=0$ ($\forall j,k$) while the $y-$components are averaged in $z-$direction $\Omega^y_{i_0+1/2,j,k+1/2}=\langle\Omega^y_j\rangle_z=\frac{1}{n_3}\Sigma_{k'} \Omega^y_{i_0+1/2,j,k'+1/2}$ ($\forall j,k$).

In summary, 3D face-centered magnetic field prescription at the Dirichlet boundary is
\begin{align}
\Omega^y_{i_0+1/2,j,k+1/2} &= 
    \begin{cases}
     0 ~ & (\bflux{})\\
     \langle\Omega^y_j\rangle_z  ~ & (\bfluxc{})
\end{cases}(\forall j,k, \text{~at internal $r-$boundary})\\
\Omega^z_{i_0+1/2,j+1/2,k} &= 0.
\end{align}

\section{Internal Static Mesh Refinement}\label{sec:smr_appendix}
The ISMR technique enables a higher $\theta$ resolution near the poles by relaxing the otherwise severely small timesteps at the poles. The technique is especially important for spinning BH cases where resolving the polar jet regions is crucial.
Motivated by the internal SMR technique in HAM-R \citep{Liska2022}, the cells are de-refined only in $\varphi$ direction in the poles to prevent small $\varphi-$edge lengths. This is performed internally to the meshblocks, over just the few rows closest to the poles.
\begin{table}[h!]
\centering
 \begin{tabular}{c c} 
 \hline
$j$ index & De-refinement Level $l$ \\ [0.5ex] 
 \hline\hline
  $j_0$ & $n$  \\
  $j_0+1$ & $n-1$ \\
  \vdots & \vdots  \\
  $j_0+k$ & $n-k$  \\
  \vdots & \vdots  \\
  $j_0+n-1$ & $1$ \\ [1ex]
 \hline
 \end{tabular}
 \caption{The $j$ indices and its de-refinement level of polar cells where the maximum de-refinement level is $n$. Here in this example, increasing $j$ index indicates further from the polar boundary and deeper into the physical cells. 
 \label{tab:ismr_setup}}
\end{table}
The de-refinement is performed such that cells closer to the pole are coarsened at a higher level as shown in Table~\ref{tab:ismr_setup}. For each de-refinement level $l$, the coarse cell contains $N_l=2^l$ fine cells in $\varphi$ direction.

De-refining conserved fluid variables $\bold{u}^{\rm fl}\equiv \sqrt{-g}(\rho u^t, T^t_t+\rho u^t, T^t_i)$ that live on cell centers is relatively straightforward. After de-refinement at a $l-$th level, the conserved fluid variables for each fine cells within a single coarse cell is $\bold{u}^{\rm fl, D}$ are simply averages $\bold{u}^{\rm fl, D}_{i,j,k_0} = \Sigma_{k=k_0}^{k_0 + N_l - 1} \bold{u}^{\rm fl}_{i,j,k}/N_l$. Hereafter, the de-refined quantities are indicated with a superscript D.

\begin{figure}[ht!]
\gridline{
\fig{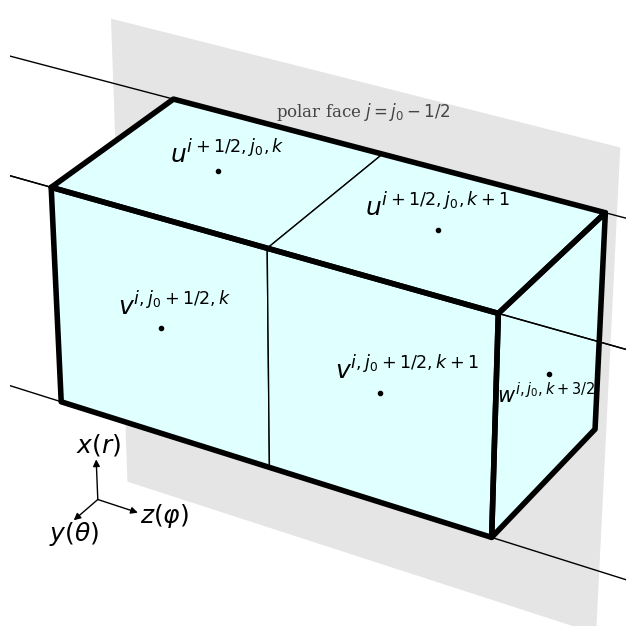}{0.5\textwidth}{}
            }   
\caption{Cell indices of the first physical layer neighboring the ghost cells when the highest level of de-refinement is $n=1$.
}\label{fig:ismr_schematic_l1}
\end{figure}

The de-refinement of the conserved magnetic fields $\sqrt{-g}B^i$ defined on face centers is less straightforward due to an extra constraint of no monopoles $\nabla \cdot \vec{B} = 0$. Following \citet{Toth2002}, the fine-grid cell spacings prior to ISMR are denoted in lower letters $\Delta x$, $\Delta y$, $\Delta z$ and the coarse-grid cell spacings are denoted in capital letters $\Delta X$, $\Delta Y$, $\Delta Z$. The fluxes of the conserved magnetic fields for the fine cells are defined as $u\equiv \sqrt{-g}B^x\Delta y \Delta z$, $v\equiv \sqrt{-g}B^y\Delta z\Delta x$, $w\equiv \sqrt{-g}B^z\Delta x \Delta y$, and the coarse cell equivalents are defined in capital letters $U$, $V$, and $W$. Then, the divergence of fine cell is
\begin{equation*}
    d_{i,j,k} = \frac{1}{\Delta x \Delta y \Delta z}(u_{i+1/2,j,k}-u_{i-1/2,j,k} + v_{i,j+1/2,k} - v_{i,j-1/2,k} + w_{i,j,k+1/2}-w_{i,j,k-1/2}),
\end{equation*}
and the constrained transport keeps each of the fine divergences $d_{i,j,k}$ to be small.

ISMR for the case of only one level of de-refinement $n=1$ is described first and is generalized below. The coarse cell is comprised of two fine cells where $\Delta X = \Delta x$, $\Delta Y = \Delta y$, $\Delta Z = 2\Delta z$. At the polar face $j=j_0-1/2$, the magnetic fields are fixed to zero $v_{i,j_0-1/2,k}=0~ (\forall \, i,k)$ due to the zero face size at the poles $(\Delta z=0)$. Then, the coarse cell's fluxes are
\begin{align*}
    U_{i+1/2,j_0,k+1/2} &= u_{i+1/2,j_0,k} + u_{i+1/2,j_0,k+1}\\
    U_{i-1/2,j_0,k+1/2} &= u_{i-1/2,j_0,k} + u_{i-1/2,j_0,k+1}\\
    V_{i,j_0+1/2,k+1/2}&= v_{i, j_0+1/2,k} + v_{i, j_0+1/2,k+1}\\
    V_{i,j_0-1/2,k+1/2}&= v_{i, j_0-1/2,k} + v_{i, j_0-1/2,k+1}=0\\
    W_{i,j_0,k+3/2}&= w_{i,j_0,k+3/2}\\
    W_{i,j_0,k-1/2}&= w_{i,j_0,k-1/2}.
\end{align*}
The resulting divergence of the restricted coarse cell can be shown to be $(d_{i,j,k} + d_{i,j,k+1})/2$ which is also small because the divergence of each fine cells $d_{i,j,k}$ is already kept small.

The new fluxes are assigned back to the two fine cells for evolving magnetic fields within the same code framework which is unaware of the coarse cells. The fine cells are treated equally as they effectively comprise a single coarse cell. The new fine cell fluxes after ISMR are
\begin{subequations}
\begin{align}\label{eq:u_fine1}
    u^{\rm D}_{i\pm1/2,j_0,k} = u^{\rm D}_{i\pm1/2,j_0,k+1} &= \frac{1}{2}U_{i\pm1/2,j_0,k+1/2},\\\label{eq:u_fine2}
    v^{\rm D}_{i,j_0-1/2,k}=v^{\rm D}_{i,j_0-1/2,k+1}&=\frac{1}{2}V_{i,j_0-1/2,k+1/2}=0,\\\label{eq:v_fine1}
    v^{\rm D}_{i,j_0+1/2,k} &= v_{i,j_0+1/2,k},\\\label{eq:v_fine2}
    v^{\rm D}_{i,j_0+1/2,k+1} &= v_{i,j_0+1/2,k+1},\\\label{eq:w_fine1}
    w^{\rm D}_{i,j_0,k+3/2} &= W_{i,j_0,k+3/2},\\\label{eq:w_fine2}
    w^{\rm D}_{i,j_0,k-1/2} &= W_{i,j_0,k-1/2}.
\end{align}
\end{subequations}
We adopt a simple first-order prolongation and divide the coarse flux into halves (Equations~\ref{eq:u_fine1}, \ref{eq:u_fine2}), except for the last 4 faces (Equations~\ref{eq:v_fine1}-\ref{eq:w_fine2}). Since $j=j_0+1/2$ face is in contact with the non-ISMR applied cells, we do not modify the $y-$fluxes in Equations~\ref{eq:v_fine1}, \ref{eq:v_fine2} in order to maintain small divergences in the neighboring non-ISMR cells. Also, for the last two $z-$fluxes (Equations~\ref{eq:w_fine1}, \ref{eq:w_fine2}), the coarse cell's $z-$face area is equivalent to the fine cells', so the coarse and fine cell fluxes are set to be equal. Note that there is one fine flux that is not yet determined, which is the $z$-face between the two fine cells, $w^{\rm D}_{i,j_0,k+1/2}$. This internal flux within a coarse cell can be used as the free parameter to keep the divergence small in both fine cells. One can derive the required internal flux to simultaneously keep divergence small for both fine cells which is
\begin{equation}\label{eq:w_central}
    w^{\rm D}_{i,j_0,k+1/2}=\frac{1}{2}(w_{i,j_0,k+3/2}+w_{i,j_0,k-1/2} + v_{i,j_0+1/2,k+1} - v_{i,j_0+1/2,k}).
\end{equation}

Next we generalize to the case of multiple levels of derefinement $n>1$. Consider a layer of cells at constant $j$ with a given de-refinement level $l\in\{1,\cdots,n\}$, where a neighboring $j+1$ layer is de-refined with a lower level $l-1$. The coarse cell dimensions are $\Delta X = \Delta x$, $\Delta Y = \Delta y$, $\Delta Z=N_l\Delta z$, where $N_l=2^l$. After performing a similar restriction and then working out the prolongation such that all fine cells have small divergence, the new fine cell fluxes are
\begin{subequations}
\begin{align}\label{eq:ul}
    u^{\rm D}_{i\pm 1/2,j,k}&=\frac{1}{N_l}\sum_{k'} u_{i\pm 1/2,j,k'}\\\label{eq:vl}
    v^{\rm D}_{i,j-1/2,k}&=\frac{1}{N_l}\sum_{k'} v_{i,j-1/2,k'}\\\label{eq:wl_L}
    w^{{\rm D}}_L &= w_{L}\\\label{eq:wl_R}
    w^{{\rm D}}_R &= w_{R}\\\label{eq:wl_C}
    w^{{\rm D}}_C&=\frac{1}{2}\left(w_L + w_R + \underbrace{\sum_{k'>k_C} v_{i,j+1/2,k'}}_\text{Right half} - \underbrace{\sum_{k'<k_C}v_{i,j+1/2,k'}}_\text{Left half}\right)\\\label{eq:wl_interpL}
    w^{\rm D}_{i,j,k-1/2} (k\in\{k_0,\cdots,k_0+N_l/2-1\})&=\frac{(N_l/2+k_0-k) w^{\rm D}_L + (k-k_0)w^{{\rm D}}_C}{N_l/2}\\\label{eq:wl_interpR}
    w^{\rm D}_{i,j,k-1/2} (k\in\{k_0+N_l/2+1,\cdots,k_0+N_l\})&=\frac{(k-k_0-N_l/2) w^{\rm D}_R + (N_l-k+k_0)w^{{\rm D}}_C}{N_l/2}
\end{align}
\end{subequations}
where $k'\in\{k_0,\cdots,k_0+N_l-1\}$ runs within a coarse cell, $k_0$ is the first fine cell within a coarse cell, $k_C=k_0+N_l/2-1/2$ is the location of the central internal face of the coarse cell. The fluxes $w_L \equiv w_{i,j,k_0-1/2}$, $w_R\equiv w_{i,j,k_0+N_l-1/2}$, and $w_C\equiv w_{i,j,k_C}$ are fluxes of the left, right, and central internal faces of the coarse cell in $\varphi$-direction. The Equations~\ref{eq:ul},~\ref{eq:vl} are simple averages of the fine cells. Note that the flux $v^{\rm D}_{i,j+1/2,k}$ is again not modified (similar to how the fluxes in Equations~\ref{eq:v_fine1},~\ref{eq:v_fine2} are unchanged) as those fluxes are de-refined by the neighboring lower-level ($l-1$) layer sharing the same face. Also, the coarse cell has the same $z-$face area so the $w_L$, $w_R$ remains unchanged in Equations~\ref{eq:wl_L},~\ref{eq:wl_R} (like Equations~\ref{eq:w_fine1}, \ref{eq:w_fine2}). The central internal face $w^{{\rm D}}_C$ exactly in the middle of the coarse cell in Equation~\ref{eq:wl_C} is set similarly to \autoref{eq:w_central} by treating left half and right half cells as two finer cells. The other $z-$fluxes are interpolations between edges $w^{\rm D}_L$, $w^{\rm D}_R$ and center of the coarse cell $w^{\rm D}_C$ (Equation~\ref{eq:wl_interpL} for left half and Equation~\ref{eq:wl_interpR} for right half cells).
A linear reconstruction is adopted only in the cells where the ISMR is applied.

\section{JKS Coordinate system}\label{sec:JKS}
\begin{figure}[ht!]
\gridline{
\fig{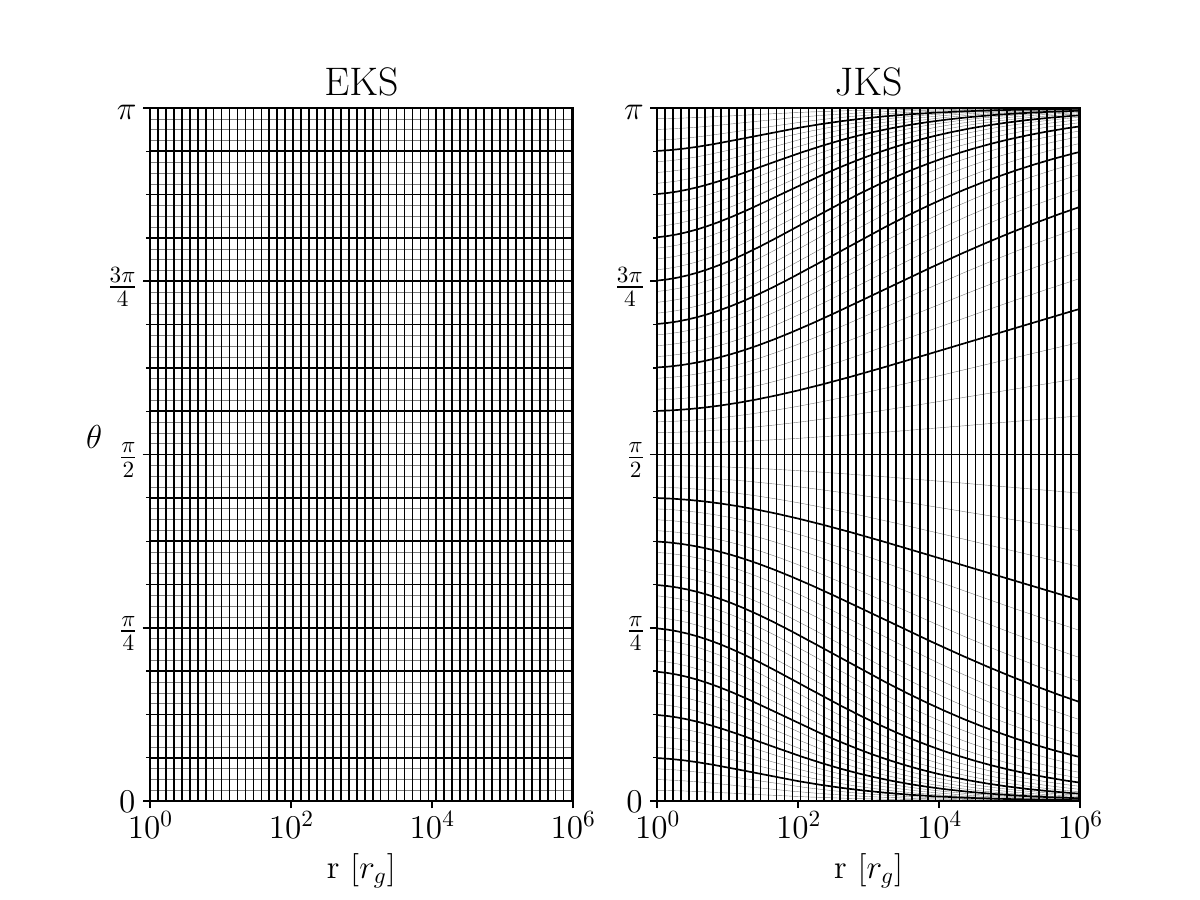}{0.7\textwidth}{}
            }   
\caption{Comparison of the EKS and JKS coordinate systems. While the $\theta-$grid is uniform at all radius for EKS, $\theta-$grid collimates near the poles at large radius for JKS.
}\label{fig:jks}
\end{figure}
The JKS coordinate system features an increasing polar resolution at increasing radius, in order to closely follow the collimation of relativistic jets and better capture the jet physics (as demonstrated in Figure~\ref{fig:n4coordcompare} in Section~\ref{sec:coord_compare}).
Figure~\ref{fig:jks} shows the comparison of the grid between EKS and JKS in $\log{r}-\theta$ plane. JKS coordinate system starts with a uniform $\theta$ spacing near the BH horizon and progressively focuses the resolution near the poles, compared to EKS's constantly uniform $\theta-$grid.

The code coordinates $x^r$, $x^\theta$, $x^\varphi$ are spaced evenly in the range $x^r\in[0, \log{(r_{\rm out})}]$, $x^\theta\in[0,1]$, $x^\varphi\in[0,2\pi]$. The $r$ and $\varphi$ are related to the code coordinates as $r(x^r) = \exp{(x^r)}$ and $\varphi(x^\varphi)=\varphi$. The $\theta$ grid is set as
\begin{equation}
    \theta(x^r, x^\theta) = \frac{\pi}{2}\left(1+ \frac{\tanh{((x^\theta -0.5)/\alpha(x^r))}}{\tanh{(0.5/\alpha(x^r))}}\right),
\end{equation}
where $\alpha(x^r)=\sigma / (x^r+0.5)$ and the parameter $\sigma$ controls the collimation level. 
Here we use the collimation $\sigma = 2$. If we assume that the jet half-opening angle is $\theta_{\rm jet}(r) = \frac{\pi}{2} r^{-0.4}$, then at a realistic Bondi radius $\sim 2\times 10^5\,r_g$, $\theta_{\rm jet}\approx 0.012$. A collimation $\sigma =2$ ensures that even at a large distance $\sim 2\times 10^5\,r_g$, the jet's half opening angle is resolved with 6 polar cells.

\section{Resolution Studies}\label{sec:resolution_study}
\begin{figure}[ht!]
\gridline{
\fig{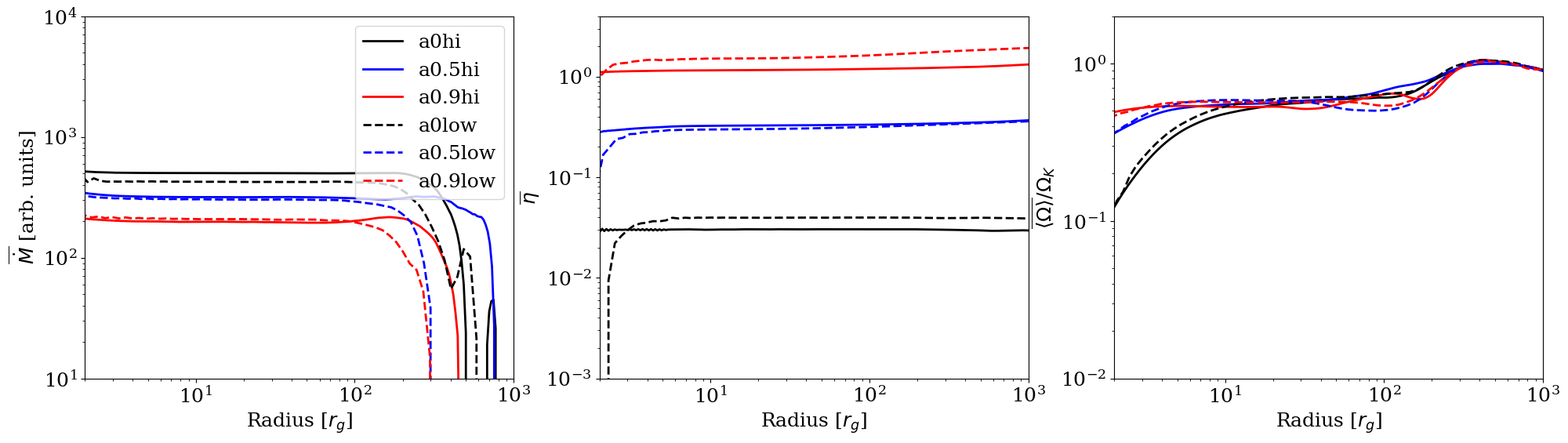}{0.99\textwidth}{}
            }   
\caption{Time-averaged profiles of accretion rate $\dot{M}$, feedback efficiency $\eta$, and shell-averaged angular velocity $\langle \Omega\rangle$ of torus simulations for different resolutions and BH spins. The runs with lower resolutions (dashed lines) show similar radial profiles as the higher resolution runs (solid lines) for three different spins (distinguished by different colors).
}\label{fig:resolution_studies}
\end{figure}
Comprehensive resolution studies \citep{White2019,Salas2024}, along with our previous work \citep{Cho2023,Cho2024}, have demonstrated that convergence is achieved even at relatively low resolutions for MAD systems. Specifically, we demonstrate that the fiducial resolution of this work is sufficient to reproduce high-resolution results for three BH spins $a_*=0$, $0.5$, and $0.9$.

Here we explore the effect of resolutions using the same set of large scale torus simulations reported in \citet{Cho2025}. In these simulations, the maximum pressure is placed at $500\,r_g$, an order of magnitude larger than the conventional tori, making them comparable in size to our small-scale simulations with Bondi radius $R_B\approx 400\,r_g$. The weakly magnetized torus runs (Section~3 in \citealt{Cho2025}) had the resolution of $288\times 192\times 144$. Since most of our small scale simulations in Section~\ref{sec:small_scale} have resolutions of $160\times 64^2$, here we simulate lower resolution versions of the large torus to approximately match our fiducial resolution. The simulations in 2 resolutions and 3 BH spins are summarized in Table~\ref{tab:res_study}. The runs \texttt{a0hi}, \texttt{a0.5hi}, and \texttt{a0.9hi} are the same set of runs (models $a0\beta100$, $a.5\beta100$, $a.9\beta100$) reported in \citet{Cho2025}. The runs \texttt{a0low}, \texttt{a0.5low}, and \texttt{a0.9low} have the same set-up except for the coarser resolution of $128\times 64^2$, which is comparable to the fiducial resolution of this work.  Both sets of runs are evolved over a total runtime of $2.8\times 10^5\,t_g$ and the time-average is taken in the range $2\times 10^5\,t_g - 2.8\times 10^5\,t_g$.
Other details of the set-up of the large torus simulations can be found in \citet{Cho2025}.

\begin{table}[h!]
\centering
 \begin{tabular}{c c c} 
 \hline
Label & BH spin $a_*$ & Resolution \\ [0.5ex] 
 \hline\hline
 \texttt{a0hi} & 0 & $288\times 192\times 144$  \\
 \texttt{a0.5hi} & 0.5 & $288\times 192\times 144$  \\
 \texttt{a0.9hi} & 0.9 & $288\times 192\times 144$  \\
 \texttt{a0low} & 0 & $128\times 64\times 64$  \\
 \texttt{a0.5low} & 0.5 & $128\times 64\times 64$  \\
 \texttt{a0.9low} & 0.9 & $128\times 64\times 64$  \\
 \hline
 \end{tabular}
 \caption{The summary of 6 torus simulations with 3 different BH spins $a_*$ and 2 different resolutions. The higher resolution runs \texttt{a0hi}, \texttt{a0.5hi}, \texttt{a0.9hi} correspond to $a0\beta100$, $a.5\beta100$, $a.9\beta100$ runs in \citet{Cho2025}. The lower resolution runs \texttt{a0low}, \texttt{a0.5low}, \texttt{a0.9low} are exactly the same but with a lower resolution comparable to the fiducial resolution in this work.
 \label{tab:res_study}}
\end{table}

The time-averaged profiles of $\dot{M}$, $\eta$, and $\langle\Omega\rangle$ are shown in Figure~\ref{fig:resolution_studies} for the 6 torus runs. Overall, there is a good agreement between the higher resolution runs (solid lines) and the lower resolution runs (dashed lines) for all three BH spin values $a_*$ (shown in different colors). Specifically, accretion rates $\dot{M}(r_H)$ are similar between high and low resolutions, also reproducing the mild dependence of accretion rates $\dot{M}$ on BH spins $a_*$. Feedback efficiency $\eta$ at lower resolution exhibits a drop near the horizon $r_H$, but the efficiency profile becomes flat starting from $\gtrsim 4\,r_g$, at which the efficiencies are extracted.

The comparison demonstrates that the fiducial resolution used here, $64$ cells in the $\theta$ and $\varphi$ directions, is sufficient to reproduce the results of high-resolution simulations for MAD accretion flows in our measurements of interest.

\bibliography{bridging_scales}{}
\bibliographystyle{aasjournal}

\end{CJK}
\end{document}